\documentclass[aps,prd,twocolumn,showpacs,superscriptaddress]{revtex4}
\usepackage{graphics}
\begin{document}
\title{Phenomenology of the $SU(3)_c\otimes SU(3)_L\otimes U(1)_X$ model 
with exotic charged leptons}

\author{Juan C. Salazar}
\affiliation{Instituto de F\'\i sica, Universidad de Antioquia,
A.A. 1226, Medell\'\i n, Colombia}
\affiliation{Depto. de F\'\i sica, Universidad de Nari\~no, A.A. 1175, Pasto, Colombia}
\author{William A. Ponce} 
\affiliation{Instituto de F\'\i sica, Universidad de Antioquia,
A.A. 1226, Medell\'\i n, Colombia}
\author{Diego A. Guti\'\i errez} 
\affiliation{Instituto de F\'\i sica, Universidad de Antioquia,
A.A. 1226, Medell\'\i n, Colombia}

\date{20 February 2007}

\begin{abstract}
A phenomenological analysis of the three-family model based on the local
gauge group $SU(3)_c\otimes SU(3)_L\otimes U(1)_X$ with exotic charged
leptons, is carried out. Instead of using the minimal scalar sector able
to break the symmetry in a proper way, we introduce an alternative set of
four Higgs scalar triplets, which combined with an anomaly-free discrete
symmetry, produce quark and charged lepton mass spectrum without 
hierarchies in the Yukawa coupling constants. We also embed the structure 
into a simple gauge group
and show some conditions to achieve a low energy gauge coupling
unification, avoiding possible conflict with proton decay bounds. By using
experimental results from the CERN-LEP, SLAC linear collider, and atomic
parity violation data, we update constraints on several parameters of the
model.
\end{abstract}


\maketitle

\section{\label{sec:intr}Introduction} 
The Standard Model (SM) based on the local gauge group $SU(3)_c\otimes
SU(2)_L\otimes U(1)_Y$ \cite{sm}, with all its successes, is in the
unaesthetic position of having no explanation of several issues such
as: hierarchical charged fermion masses, fermion mixing angles, charge quantization, strong CP
violation, replication of families, neutrino masses and oscillations~\cite{valle}, etc.. 
All this make us think that we must call for extensions of the model.

Doing physics beyond the SM may imply to introduce a variety of new ingredients such as extra fermion
fields (adding a right-handed neutrino field to each family constitute its
simplest extension and has several consequences, as the implementation of
the see-saw mechanism for the neutrinos, and the enlarging of the possible number of local
Abelian symmetries that can be gauged simultaneously). Also one may
include standard and nonstandard new scalar field representations with and without Vacuum Expectation Values (VEV), and extra gauge bosons which imply an enlarging of the local gauge group. Discrete symmetries and supersymmetry (SUSY) are also common extensions of the SM~\cite{moha}.

Interesting extensions of the SM are based on the local gauge group~\cite{pf, vl, ozer, sher, pfs, pgs} 
$SU(3)_c\otimes SU(3)_L\otimes U(1)_X$ (called hereafter 3-3-1 for short). The several possible structures
enlarge the SM in its gauge, scalar, and fermion sectors. Let us mention some outstanding features of 3-3-1 models: they are free of gauge anomalies if and only if the number of families is a multiple of three~\cite{pf, vl, ozer}; a Peccei-Quinn chiral symmetry can be implemented easily~\cite{pq, pal}; the fact that one quark family has different quantum numbers than the other two may be used to explain the heavy top quark mass~\cite{ff, canal}; the scalar sector includes several good candidates for dark matter~\cite{sanchez}, the lepton content is suitable for explaining some neutrino properties~\cite{kita1, kita2}, and last but not least, the hierarchy in the Yukawa coupling constants can be avoided by implementing several universal see-saw mechanisms~\cite{canal, seesaw, dwl}.

So far, there are in the literature studies of five different 3-3-1 lepton flavor structures for three families, belonging to two different electric charge embedding into $SU(3)_L\otimes U(1)_X$, being the most popular one the original Pisano-Pleitez-Frampton model~\cite{pf} (called the minimall model)in which the three left-handed lepton
components for each family in the SM are associated to three $SU(3)_L$ triplets as
$(\nu_l,l^-,l^+)_L$, where $l=e,\;\mu,\;\tau$ is a family index for the
lepton sector, $\nu_l$ stands for the neutrino related to the flavor $l$,
and $l^+_L$ is the right-handed isospin singlet of the charged lepton
$l^-_L$. 

In a different embedding of the electric charge operator, the three left-handed lepton triplets are of the form $(\nu_l, l^-, \nu_l^c)_L$, $l=e,\;\mu,\;\tau$; where $\nu_l^c$ is related to the right-handed component of the neutrino field (a model with ``right-handed neutrinos"~\cite{vl}), with $l_L^+$ becoming three $SU(3)_L$ singlets. For the same charge embedding, an almost unknown alternative of a 3-3-1 fermion structure is provided in Ref.~\cite{ozer}, in which the three $SU(3)_L$ lepton triplets are of the form
$(\nu_l, l^-, E_l^-)_L$, $l=e,\;\mu,\;\tau$;  where $E_l^-$ stands for an exotic charged lepton per family, with $l_L^+$ and $E_{lL}^+$ being six $SU(3)_L$ singlets (a model with ``exotic charged leptons").

Contrary to the former three structures in which each lepton generation is treated identically, two more new models are analyzed in Ref.~\cite{sher}, which are characterized by each lepton generation having a different representation under the gauge group. Even further, more possible 3-3-1 fermion structures can be found in Refs.~\cite{pfs, pgs}, where also a classification of all the models without exotic electric charges is presented (if exotic electric charges are allowed the number of models run to infinity).

The aim of this paper is to find, for the version of the model that includes ``exotic charged leptons"\cite{ozer}, the minimal set of ingredients able to implement universal see-saw mechanisms in the three charged fermion sectors, with the analysis done in a similar way that the one presented in Refs.~\cite{canal, dwl}, where a related calculation was carried through for the model with `` right-handed neutrinos" (model that, contrary to the present one, does not contain exotic electrons, becoming thus unable by itself to generate see-saw masses for charged leptons~\cite{dwl}). It will be shown in what follows that a convenient set of scalar fields, combined with a discrete symmetry, produces an appealing fermion mass spectrum without hierarchies for the Yukawa coupling constants. Besides, we are also going to study the embedding and unification of this structure into $SU(6)$, and set updated constraints on
several parameters of the model.

This paper is organized as follows: in Sec.~\ref{sec:sec2} we review the
model, introduce the new scalar sector, embed the structure into a
covering group and calculate the charged and neutral electroweak currents;
in Sec.~\ref{sec:sec3} we study the charged fermion mass spectrum; in
Sec.~\ref{sec:sec4} we do the renormalization group equation analysis and
show the conditions for the gauge coupling unification; in
Sec.~\ref{sec:sec5} we constraint several parameters of the model by fixing new bounds on the mixing angle between the
two flavor diagonal neutral currents present in the model, and finally, in
Sec.~\ref{sec:sec6}, we present our conclusions.

%%%%%%%%%%%%%%%%%%%%%%%%%%%%%%%%%%%%%%%%%%%%%%%%%%%%%%%%%%%%

\section{\label{sec:sec2}The model} 
The model we are about to study here was introduced in the literature for
the first time in Ref.~\cite{ozer}. Some of the formulas quoted in
subsections \ref{sec:sec21} and \ref{sec:sec26} of this section, are taken
from Refs.¿\cite{ozer} and¿\cite{pgs}. Corrections to some minor
printing mistakes in the original papers are included.  
\subsection{\label{sec:sec21}The Gauge Group} As it was stated above, the
model we are interested in, is based on the local gauge group
$SU(3)_c\otimes SU(3)_L\otimes U(1)_X$ which has 17 gauge bosons: one
gauge field $B^\mu$ associated with $U(1)_X$, the 8 gluon fields $G^\mu$
associated with $SU(3)_c$ which remain massless after spontaneous breaking of the electroweak 
symmetry, and another 8 gauge fields associated with $SU(3)_L$ that we
write for convenience as \cite{pgs}
\begin{equation}\label{maga}
\sum_{\alpha=1}^8\lambda^\alpha A^\mu_\alpha=\sqrt{2}\left(
\begin{array}{ccc}D^\mu_1 & W^{+\mu} & K^{+\mu} \\ W^{-\mu} & D^\mu_2 &
K^{0\mu} \\ K^{-\mu} & \bar{K}^{0\mu} & D^\mu_3 \end{array}\right), 
\end{equation}
where $D^\mu_1=A_3^\mu/\sqrt{2}+A_8^\mu/\sqrt{6},\;
D^\mu_2=-A_3^\mu/\sqrt{2}+A_8^\mu/\sqrt{6}$, and
$D^\mu_3=-2A_8^\mu/\sqrt{6}$. $\lambda_\alpha, \; \alpha=1,2,...,8$, are the eight
Gell-Mann matrices normalized as $Tr(\lambda^\alpha\lambda^\beta)  
=2\delta_{\alpha\beta}$.

The charge operator associated with the unbroken gauge symmetry $U(1)_Q$ 
is given by:
\begin{equation}\label{chargo}
Q=\frac{\lambda_{3L}}{2}+\frac{\lambda_{8L}}{2\sqrt{3}}+XI_3
\end{equation}
where $I_3=Diag.(1,1,1)$ is the diagonal $3\times 3$ unit matrix, and the 
$X$ values are related to the $U(1)_X$ hypercharge and are fixed by 
anomaly cancellation. 
The sine square of the electroweak mixing angle is given by 
\begin{equation}\label{ewk}
S_W^2=3g_1^2/(3g_3^2+4g_1^2)
\end{equation}
where $g_1$ and $g_3$ are the coupling 
constants of $U(1)_X$ and $SU(3)_L$ respectively, and the photon field is 
given by~\cite{ozer, pgs} 
\begin{equation}\label{foton}
A_0^\mu=S_WA_3^\mu+C_W\left[\frac{T_W}{\sqrt{3}}A_8^\mu + 
\sqrt{(1-T_W^2/3)}B^\mu\right],
\end{equation}
where $C_W$ and $T_W$ are the cosine and tangent of the electroweak mixing 
angle, respectively. 

There are two weak neutral currents in the model 
associated with the two flavor diagonal neutral gauge weak bosons 
\begin{eqnarray}\nonumber \label{zzs}
Z_0^\mu&=&C_WA_3^\mu-S_W\left[\frac{T_W}{\sqrt{3}}A_8^\mu + 
\sqrt{(1-T_W^2/3)}B^\mu\right], \\ \label{zetas}
Z_0^{\prime\mu}&=&-\sqrt{(1-T_W^2/3)}A_8^\mu+\frac{T_W}{\sqrt{3}}B^\mu,
\end{eqnarray}
and another electrically neutral current associated with the 
gauge boson $K^{0\mu}$. In the former expressions 
$Z^\mu_0$ coincides with the weak neutral current of the SM¿\cite{ozer, pgs}. Using  
Eqs. (\ref{foton}) and (\ref{zetas}) we can read the gauge boson $Y^\mu$ 
associated with the $U(1)_Y$ hypercharge of the SM 
\begin{equation} \label{hyper}
Y^\mu=\left[\frac{T_W}{\sqrt{3}}A_8^\mu + 
\sqrt{(1-T_W^2/3)}B^\mu\right].
\end{equation}

Equations (\ref{maga}-\ref{hyper}) presented here are common to all the 3-3-1 gauge structures without exotic electric charges~\cite{vl, ozer, sher} as it is analyzed in Refs.~\cite{pfs, pgs}.

\subsection{\label{sec:sec22}The spin 1/2 particle content} 
The quark content for the three families is the following~\cite{ozer}:  
$Q^i_{L}=(d^i,u^i,U^i)_L^T\sim(3,3^*,1/3),\;i=1,2,$ for two families,
where $U^i_L$ are two exotic quarks of electric charge $2/3$ (the numbers
inside the parenthesis stand for the $[SU(3)_c,SU(3)_L,U(1)_X]$ quantum
numbers in that order); $Q^3_{L}=(u^3,d^3,D)^T_L\sim (3,3,0)$, where $D_L$
is an exotic quark of electric charge $-1/3$. The right handed quarks are
$u^{ac}_{L}\sim (3^*,1,-2/3),\; d^{ac}_{L}\sim (3^*,1,1/3)$ with
$a=1,2,3,$ a family index, $U^{ic}_{L}\sim (3^*,1,-2/3),\;i=1,2$, and
$D^c_L\sim (3^*,1,1/3)$.

The lepton content is given by~\cite{ozer} three $SU(3)_L$ triplets $L_{lL} =(\nu_l^0, l^-,E_l^-)^T_L\sim (1,3,-2/3)$, for $l=e,\mu,\tau$ a lepton family index,
and $\nu^0_l$ the neutrino field associated to the flavor $l$. The six lepton singlets are $l^+_{L}\sim(1,1,1)$, and $E^+_{lL}\sim(1,1,1)$. Notice in this model the presence of three exotic electrons $E_l^-$ of electric charge $-1$ (used in what follows to implement the universal see-saw mechanism in the charged lepton sector), and the fact that it does not contain right-handed neutrinos. For this model, universality for the known leptons of the three families is present at tree level in the weak basis.

With the former quantum numbers it is just a matter of counting to check
that the model is free of the following gauge anomalies\cite{pfs}:  
$[SU(3)_c]^3$; (as in the SM $SU(3)_c$ is vectorlike); $[SU(3)_L]^3$ (six
triplets and six anti-triplets), $[SU(3)_c]^2U(1)_X; \; [SU(3)_L]^2U(1)_X
; \;[grav]^2U(1)_X$ and $[U(1)_X]^3$, where $[grav]^2U(1)_X$ stands for
the gravitational anomaly as described in Ref.~\cite{del}.

\subsection{\label{sec:sec23}The new scalar sector}
Instead of using the set of three triplets of Higgs scalars introduced in the original paper  \cite{ozer}, or the most economical set of two triplets introduced in Ref. \cite{pgs} (none of them able to produce a realistic mass spectrum), we propose here to start working with the following set of four Higgs scalar fields, and VEV:
\begin{eqnarray}\label{higgsses} \nonumber
\langle\phi_1^T\rangle &=&\langle(\phi^+_1, \phi^0_1,\phi^{'0}_1)\rangle = 
\langle(0,0,v_1)\rangle \sim (1,3,1/3) \\ \nonumber
\langle\phi_2^T\rangle &=&\langle(\phi^+_2, \phi^0_2,\phi^{'0}_2)\rangle = 
\langle(0,v_2,0)\rangle \sim (1,3,1/3) \\ \nonumber
\langle\phi_3^T\rangle &=&\langle(\phi^0_3, \phi^-_3,\phi^{'-}_3)\rangle = 
\langle(v_3,0,0)\rangle \sim (1,3,-2/3) \\ \nonumber
\langle\phi_4^T\rangle &=&\langle(\phi^+_4, \phi^0_4,\phi^{'0}_4)\rangle = 
\langle(0,0,V)\rangle \sim (1,3,1/3), \\
\end{eqnarray}
with the hierarchy $v_1\sim v_2\sim v_3\sim v\sim 10^2$ GeV $<< V$. Notice that the vacuum has been aligned arbitrarily such that $\langle\phi_1^0\rangle=\langle\phi_2^{\prime 0}\rangle= \langle\phi_4^0\rangle=0$, in order to accomplish for the following facts:
\begin{itemize}
\item To have at the electroweak scale $v$ an effective theory with properties resembling the two Higgs doublet extension of the SM.
\item To properly implement several universal see-saw mechanisms~\cite{seesaw}.
\item To avoid unnecessary mixing in the electroweak gauge boson sector~\cite{pgs}.
\end{itemize}

The alternative of minimizing the scalar potential is a complicated and fruitless task at this stage of development of this particular model.

The set of scalars and VEV in Eq.~(\ref{higgsses}) break the 
$SU(3)_c\otimes SU(3)_L\otimes U(1)_X$ symmetry in two steps, 
\begin{eqnarray*}
SU(3)_c\otimes SU(3)_L\otimes U(1)_X &\stackrel{(V+v_1)}{\longrightarrow}& \\
SU(3)_c\otimes SU(2)_L\otimes U(1)_Y &\stackrel{(v_2+v_3)}{\longrightarrow}& SU(3)_c\otimes U(1)_Q,
\end{eqnarray*}
which allows for the matching conditions $g_2=g_3$ and 
\begin{equation}\label{mc}
\frac{1}{g^{2}_Y}=\frac{1}{g_1^2}+\frac{1}{3g_2^2},
\end{equation}
where $g_2$ and $g_Y$ are the gauge coupling constants of 
the $SU(2)_L$ and $U(1)_Y$ gauge groups in the SM, respectively. 

We will see in what follows that this scalar structure 
properly breaks the symmetry, provides with masses for the gauge 
bosons and, combined with a discrete symmetry, it is enough to produce a 
consistent mass spectrum for the charged fermion sectors (quarks and leptons). The mass spectrum for the neutral lepton sector requires new ingredients as it is going to be analyzed in Sec.~\ref{sec:sec34}.

\subsection{\label{sec:sec24} $SU(6)\supset SU(5)$ as a covering group} 
The Lie algebra of $SU(3)\otimes SU(3)\otimes U(1)$ is a maximal  
subalgebra of the simple algebra of $SU(6)$. The five fundamental 
irreducible representations (irreps) of $SU(6)$ are: 
$\{6\},\{6^*\},\{15\},\{15^*\}$ and the \{20\} which is real. The branching rules for 
these fundamental irreps into 
$SU(3)_c\otimes SU(3)_L\otimes U(1)_X$ are \cite{slansky}:
\begin{eqnarray}\label{branching}\nonumber
\{6\} &\rightarrow &(3,1,-1/3)\oplus (1,3,1/3), \\ \nonumber
\{15\} &\rightarrow & (3^*,1,-2/3)\oplus (1,3^*,2/3) \oplus (3,3,0), \\ \nonumber
\{20\} &\rightarrow & (1,1,1)\oplus (1,1,-1)\oplus (3,3^*,1/3)\\ \nonumber 
       & & \oplus (3^*,3,-1/3),
\end{eqnarray}
where we have normalized the $U(1)_X$ hypercharge according to our convenience.

From these branching rules and from the fermion structure presented in Sec.~\ref{sec:sec22}, it is clear that all the particles in the 3-3-1 model with exotic electrons, can 
be included in the following $SU(6)$ reducible representation
\begin{equation}\label{reducible}
4\{6^*\}+6\{20\}+5\{15\}+3\{15^*\},
\end{equation}
which, besides the particles in the representations already stated 
in the previous section, includes new exotic particles, as for example 
\begin{eqnarray*}
(N^0,E^+,E^{\prime +})_L&\sim&(1,3^*,2/3)\subset \{15\}, \\
E^-_L&\sim&(1,1,-1)\subset \{20\}, \\
(D^{\prime c},U^{\prime c},U^{\prime\prime c})_L&\sim&(3^*,3,-1/3)\subset 
\{20\}. \end{eqnarray*}

The analysis reveals that the reducible representation in
(\ref{reducible}) is free of anomalies (irrep $\{20\}$ is real and
anomaly-free, and the anomaly of one $\{15\}$ is twice the anomaly of a
$\{6\}$~\cite{slansky}).

It is clear from the following decomposition of irrep $\{6^*\}$ of $SU(6)$
into $SU(5)\otimes U(1)$
\begin{equation}
\{6^*\}=\{d^c,N_E^0,E^-,N^{0c}_E\}_L\longrightarrow \{d^c,N^0_E,E^-\}_L\oplus N^{0c}_{EL}, 
\end{equation}
that for $N^0_{EL}=\nu_{eL}$ and $E^-_L=e^-_L$, we obtain the known $SU(5)$ model of Georgi and Glashow \cite{gg}; so in some sense, the $SU(6)$ here is an extension of one of the first Grand Unified Theories (GUT) studied in the literature.

\subsection{\label{sec:sec25}The gauge boson sector}
After breaking the symmetry with $\langle\phi_i\rangle,\; i=1,\dots ,4$, and
using the covariant derivative for triplets $D^\mu=\partial^\mu
-ig_3\lambda^{\alpha}_LA^\mu_\alpha/2-ig_1XB_\mu I_3$, we get the 
following mass terms in the gauge boson sector.

\subsubsection{\label{sec:sec251}Spectrum in the charged gauge boson sector}
A straightforward calculation shows that the charged gauge bosons
$K^\pm_\mu$ and $W^\pm_\mu$ do not mix with each other and get the
following masses: $M^2_{K^\pm}=g_3^2(V^2+v_1^2+v_3^2)/2$ and
$M_W^2=g_3^2(v_2^2+v_3^2)/2$, which for $g_3=g_2$ and using the
experimental value $M_{W}=80.423 \pm 0.039$ GeV (experimental values
throughout the paper are taken from Ref.~\cite{pdb}) implies
$\sqrt{v_2^2+v_3^2}\simeq 175$ GeV. In the same way $K^{0\mu}$ (and its
antiparticle $\bar{K}^{0\mu}$) does not mix with the other two
electrically neutral gauge bosons and gets a bare mass
$M_{K^0}^2=g_3^2(V^2+v_1^2+v_2^2)/2\approx M^2_{K^\pm}$. Notice that $v_1$
does not contribute to the $W^\pm$ mass because it is associated with an
$SU(2)_L$ singlet Higgs scalar.

\subsubsection{\label{sec:sec252}Spectrum in the neutral gauge boson sector}
The algebra now shows that in this sector the photon field $A_0^\mu$ in Eq. (\ref{foton}) decouples from $Z_0^\mu$ and $Z_0^{\prime\mu}$ and remains massless. Then in the basis $(Z_0^\mu,Z_0^{\prime\mu})$ we obtain the following $2\times 2$ mass matrix 
\begin{equation} \label{gauges}
\frac{\eta^2g_3^2}{4C_W^2}\left( \begin{array}{cc}
\frac{v_2^2+v_3^2}{\eta^2}& 
\frac{v_2^2C_{2W}-v_3^2}{\eta} \\
\frac{v_2^2C_{2W}-v_3^2}{\eta} & 
v_2^2C^2_{2W}+ v_3^2+4(V^2+v_1^2)C_W^4
\end{array}\right),
\end{equation}
where $C_{2W}=C_W^2-S_W^2$ and $\eta^{-2}=(3-4S_W^2)$. This matrix provides with a mixing between $Z^\mu_0$ and $Z^{\prime\mu}_0$ of the form 
\begin{eqnarray} \label{tan} \nonumber
\tan(2\theta)&= \frac{2\sqrt{(3 - 4S^2_W)}(v_2^2C_{2W}-v_3^2)}
{4C_W^4(V^2+v_1^2) - 2v_3^2C_{2W}-v_2^2(3-4S_W^2-C_{2W}^2)}\\ 
&\stackrel{V\rightarrow\infty}{\longrightarrow}0 .
\end{eqnarray}
The physical fields are then 

\begin{eqnarray}\nonumber
Z_1^\mu&=&Z^\mu_0 
\cos\theta-Z^{\prime\mu}_0\sin\theta \; ,\\ \nonumber
Z_2^\mu&=&Z^\mu_0 \sin\theta+Z'^\mu_0 \cos\theta .
\end{eqnarray} 
An updated bound on the mixing angle $\theta$ is going to be calculated in 
Sec.~\ref{sec:sec5} using experimental results.

\subsection{\label{sec:sec26}Currents}

\subsubsection{\label{sec:sec261}Charged currents}
The Hamiltonian for the currents charged under the generators of 
$SU(3)_L$ is  :
\[H^{CC}=g_3(W^+_\mu J^\mu_{W^+}+K^+_\mu J_{K^+}^\mu+K^0_\mu 
J_{K^0}^\mu)/\sqrt{2}+H.c.,\] 
where 
\begin{eqnarray}\nonumber
J_{W^+}^\mu&=& \left[\bar{u}^3_{L}\gamma^\mu d^3_{L} 
-(\sum_{i=1}^2\bar{u}^i_{L}\gamma^\mu d^i_{L})
+\sum_{l=e,\mu,\tau}
\bar{\nu}_{l L}\gamma^\mu l^-_{L}\right]\\ \nonumber 
J^\mu_{K^+}&=&\left[(\sum_{i=1}^2\bar{U}^i_{L}\gamma^\mu 
d^i_{L})-\bar{u}^3_{L}\gamma^\mu 
D_{L} + \sum_{l=e,\mu,\tau}
\bar{\nu}^0_{l L}\gamma^\mu E^-_{lL}\right] \\ \nonumber 
J^\mu_{K^0}&=&\left[(\sum_{i=1}^2\bar{u}^i_{L}\gamma^\mu 
U^i_{L})-\bar{D}_{L}\gamma^\mu d^3_{L} + \sum_{l=e,\mu,\tau}
\bar{E}^-_{l L}\gamma^\mu l^-_{L} \right], \nonumber
\end{eqnarray} 
where $K^0_\mu$ is an electrically neutral gauge boson, but it carries a
kind of weak V-isospin charge, besides it is flavor nondiagonal.

\subsubsection{\label{sec:sec262}Neutral currents}
The neutral currents $J_\mu(EM), \; J_\mu(Z)$ and $J_\mu(Z^\prime)$
associated with the Hamiltonian $H^0 = eA^\mu J_\mu(EM)+(g_3/{C_W})Z^\mu
J_\mu(Z)+ (g_1/\sqrt{3})Z^{\prime\mu} J_\mu(Z^\prime)$ are
\begin{eqnarray}\nonumber
J_\mu(EM)&=&{2\over 3}\left[\sum_{a=1}^3\bar{u}_a\gamma_\mu u_a +
\sum_{i=1}^2\bar{U}^i\gamma_\mu U^i \right] \\ \nonumber
&-& {1\over 3}\left[\sum_{a=1}^3\bar{d}^a\gamma_\mu d^a+ 
\bar{D}\gamma_\mu D \right]  \\ \nonumber
&-& \sum_{l=e,\mu,\tau}(\bar{l}^-\gamma_\mu l^- + 
\bar{E}^-_l\gamma_\mu E^-_l) \\ \nonumber
=\sum_f q_f\bar{f}\gamma_\mu f,\\* \nonumber
J_\mu(Z)&=&J_{\mu,L}(Z)-S^2_WJ_\mu(EM),\\ \nonumber
J_\mu(Z^\prime)&=& J_{\mu,L}(Z^\prime)+T_WJ_\mu(EM), 
\end{eqnarray}
where $e=g_3S_W=g_1C_W\sqrt{(1-T_W^2/3)}>0$ is the unit of electric charge, 
$q_f$ is the electric charge of the fermion $f$ in units of $e$, and 
$J_\mu(EM)$ is the electromagnetic current. 

The left-handed currents are
\begin{eqnarray}\label{jz1} \nonumber
J_{\mu,L}(Z)&=&{1\over 2}[\sum_{a=1}^3(\bar{u}^a_{L}\gamma_\mu u^a_{L}
-\bar{d}^a_{L}\gamma_\mu d^a_{L})\\ \nonumber
&+& \sum_{l=e,\mu,\tau}(\bar{\nu}_{l L}\gamma_\mu \nu_{l L} 
-\bar{l}^-_{ L}\gamma_\mu l^-_{l L})] \\ 
&=&\sum_F \bar{F}_LT_{3f}\gamma_\mu F_L ,
\end{eqnarray}
\begin{eqnarray}\label{jz2}\nonumber
J_{\mu,L}(Z^\prime)&=& 
S^{-1}_{2W}[(\bar{u}_{1L}\gamma_\mu u_{1L}+\bar{u}_{2L}\gamma_\mu u_{2L}-
\bar{d}_{3L}\gamma_\mu d_{3L}\\ \nonumber
&-&\sum_l(\bar{\nu}_{lL}\gamma_\mu \nu_{lL})] \\  \nonumber
&+&T^{-1}_{2W}[(\bar{d}_{1L}\gamma_\mu d_{1L}+\bar{d}_{2L}\gamma_\mu 
d_{2L}-\bar{u}_{3L}\gamma_\mu u_{3L}\\ 
\nonumber
&-&\sum_l(\bar{l}^-_{L}\gamma_\mu l^-_{L})] \\ \nonumber
&-&T^{-1}_{W}[(\bar{U}_{1L}\gamma_\mu U_{1L}+
\bar{U}_{2L}\gamma_\mu U_{2L}-
\bar{D}_{L}\gamma_\mu D_{L}\\
&-&\sum_l(\bar{E}^-_{lL}\gamma_\mu E^-_{lL})] 
=\sum_F\bar{F}_L T^\prime_{3f}\gamma_\mu F_L,
\end{eqnarray}
where $S_{2W}=2S_WC_W,\; T_{2W}=S_{2W}/C_{2W}, \;T_{3f}=Dg(1/2,-1/2,0)$ is
the third component of the weak isospin, $T^\prime_{3f}=Dg(S^{-1}_{2W},
T^{-1}_{2W}, -T_W^{-1})$ is a convenient $3\times 3$ diagonal matrix,
acting both of them on the representation 3 of $SU(3)_L$ (the negative
value when acting on the representation $3^*$, which is also true for 
the matrix $T_{3f}$) and $F$ is a generic
symbol for the representations 3 and $3^*$ of $SU(3)_L$.  Notice that
$J_\mu(Z)$ is just the generalization of the neutral current present in
the SM. This allows us to identify $Z_\mu$ as the neutral gauge boson of
the SM, which is consistent with Eqs.(\ref{zzs}) and (\ref{hyper}).

The couplings of the flavor diagonal mass eigenstates $Z_1^\mu$ and 
$Z_2^\mu$ are given by:
\begin{eqnarray} \nonumber
H^{NC}&=&\frac{g_3}{2C_W}\sum_{i=1}^2Z_i^\mu\sum_f\{\bar{f}\gamma_\mu
[a_{iL}(f)(1-\gamma_5)\\ \nonumber & & 
+a_{iR}(f)(1+\gamma_5)]f\} \\ \nonumber
      &=&\frac{g_3}{2C_W}\sum_{i=1}^2Z_i^\mu\sum_f\{\bar{f}\gamma_\mu
      [g(f)_{iV}-g(f)_{iA}\gamma_5]f\},
\end{eqnarray}
where
\begin{eqnarray} \nonumber
a_{1L}(f)&=&\cos\theta(T_{3f}-q_fS^2_W) 
+\Theta\sin\theta (T^\prime_{3f}-q_fT_W), \\ \nonumber
a_{1R}(f)&=&-q_f\left(\cos\theta S_W^2 + \Theta\sin\theta T_W\right),\\ \nonumber
a_{2L}(f)&=&\sin\theta (T_{3f}-q_fS^2_W)-\Theta\cos\theta (T^\prime_{3f}-q_fT_W),\\ 
\label{a}
a_{2R}(f)&=&-q_f\left(\sin\theta S^2_W-\Theta\cos\theta T_W\right),
\end{eqnarray}
where $\Theta = S_WC_W/\sqrt{(3-4S_W^2)}$. From this coefficients we can read 
\begin{eqnarray} \nonumber
g(f)_{1V}&=&\cos\theta (T_{3f}-2q_fS^2_W) + \Theta\sin\theta (T^\prime_{3f}-2q_fT_W), \\ \nonumber
g(f)_{2V}&=&\sin\theta (T_{3f}-2q_fS^2_W)-\Theta\cos\theta (T^\prime_{3f}-2q_fT_W),\\ 
g(f)_{1A}&=&\cos\theta T_{3f}+\Theta\sin\theta T^\prime_{3f}, \\ \nonumber \label{g}
g(f)_{2A}&=&\sin\theta T_{3f}-\Theta\cos\theta T^\prime_{3f}.
\end{eqnarray}
The values of $g_{iV},\; g_{iA}$ with $i=1,2$ are listed in Tables \ref{tab1} 
and \ref{tab2}.

\begin{table*}
\caption{\label{tab1}The $Z_1^\mu\longrightarrow \bar{f}f$ couplings.}
\begin{ruledtabular}
\begin{tabular}{lcc}
$f$ & $g(f)_{1V}$ & $g(f)_{1A}$ \\ \hline

$u^{1,2}$&$({1\over 2}-{4S_W^2 \over 3})\cos\theta -  
\Theta (T_{2W}^{-1}+{4T_W\over 3})\sin\theta$ & 

${1\over 2}\cos\theta - \Theta T_{2W}^{-1}\sin\theta$\\

$u^{3}$& 
$({1\over 2}-{4S_W^2 \over 3})\cos\theta 
+\Theta(s_{2W}^{-1}-{4T_W \over 3})\sin\theta$
& ${1\over 2}\cos\theta + \Theta S_{2W}^{-1}\sin\theta$  \\ 

$d^{1,2}$ & $(-{1\over 2}+{2S_W^2\over 3})\cos\theta 
-\Theta(S_{2W}^{-1}-{2T_W\over 3})\sin\theta$ & 

$-{1\over 2}\cos\theta - \Theta S_{2W}^{-1}\sin\theta$\\

$d^{3}$ & $(-{1\over 2}+{2S_W^2\over 3})\cos\theta 
+\Theta(T_{2W}^{-1}+{2T_W\over 3})\sin\theta$ 

& $-{1\over 2}\cos\theta +\Theta T_{2W}^{-1}\sin\theta$ \\

$U^{1,2}$ & $-{4S_W^2\over 3}\cos\theta 
+\Theta (T_W^{-1}-{4T_W\over 3})\sin\theta $ &
$\Theta T_W^{-1}\sin\theta $ \\

$D $ & ${2S_W^2\over 3}\cos\theta -\Theta(T_W^{-1}-
{2T_W\over 3})\sin\theta $ & $-\Theta T_W^{-1}\sin\theta $ \\

$e^-,\; \mu^-, \; \tau^-$& $(-{1\over 2}+2S_W^2)\cos\theta +
\Theta(T_{2W}^{-1}+2T_W)\sin\theta $ 
& $ -{1\over 2}\cos\theta +\Theta T_{2W}^{-1}\sin\theta $\\

$\nu_e, \; \nu_\mu, \; \nu_\tau$ & ${1\over 2}\cos\theta +\Theta 
S_{2W}^{-1}\sin\theta$ 
& ${1\over 2}\cos\theta +\Theta S_{2W}^{-1}\sin\theta$ \\

$E^-_e, \; E^-_\mu, \; E^-_\tau$ & $2S_W^2\cos\theta -
\Theta (T_{W}^{-1}-2T_W)\sin\theta $ &
$-\Theta T_{W}^{-1}\sin\theta $ \\ 

\end{tabular}
\end{ruledtabular}
\end{table*}

\begin{table*}
\caption{\label{tab2}The $Z_2^\mu\longrightarrow \bar{f}f$ couplings.}
\begin{ruledtabular}
\begin{tabular}{lcc}
$f$ & $g(f)_{2V}$ & $g(f)_{2A}$ \\ \hline

$u^{1,2}$&$({1\over 2}-{4S_W^2 \over 3})\sin\theta +\Theta (T_{2W}^{-1}+
{4T_W\over 3})\cos\theta $ & ${1\over 2}\sin\theta +\Theta
T_{2W}^{-1}\cos\theta $\\

$u^{3}$& $({1\over 2}-{4S_W^2 \over 3})\sin\theta -\Theta (S_{2W}^{-1}-
{4T_W\over 3})\cos\theta $
& ${1\over 2}\sin\theta -\Theta S_{2W}^{-1}\cos\theta $ \\ 

$d^{1,2}$ & $(-{1\over 2}+{2S_W^2\over 3})\sin\theta +
\Theta (S_{2W}^{-1}-{2T_W\over 3})\cos\theta $ & 
$-{1\over 2}\sin\theta +\Theta S_{2W}^{-1}\cos\theta $\\

$d^{3}$ & $(-{1\over 2}+{2S_W^2\over 3})\sin\theta-
\Theta(T_{2W}^{-1}+{2T_W\over 3})\cos\theta $
& $-{1\over 2}\sin\theta -\Theta T_{2W}^{-1}\cos\theta $ \\

$U^{1,2}$ & ${-4S_W^2\over 3}\sin\theta 
-\Theta(T_W^{-1}-{4T_W\over 3})\cos\theta $ &
$-\Theta T_W^{-1}\cos\theta $ \\

$D$ & ${2S_W^2\over 3}\sin\theta + 
\Theta(T_W^{-1}-{2T_W\over 3})\cos\theta $ & 
$\Theta T_W^{-1}\cos\theta $\\

$e^-, \; \mu^-,\; \tau^-$& $(-{1\over 2}+2S_W^2)\sin\theta
-\Theta(T_{2W}^{-1}+ 2T_W)\cos\theta $ & 
$-{1\over 2}\sin\theta -\Theta T_{2W}^{-1}\cos\theta $\\

$\nu_e, \; \nu_\mu, \; \nu_\tau $ & ${1\over 2}\sin\theta -\Theta S_{2W}^{-1}\cos\theta $ & 
${1\over 2}\sin\theta - \Theta S_{2W}^{-1}\cos\theta $ \\

$E^-_e, \; E^-_\mu, \; E^-_\tau$ & $2S_W^2\sin\theta +\Theta
(T_{W}^{-1}-2T_W)\cos\theta $ & $\Theta
T_W^{-1}\cos\theta $ \\

\end{tabular}
\end{ruledtabular}
\end{table*}

As can be seen, in the limit $\theta = 0$ the couplings of
$Z_1^\mu$ to the ordinary leptons and quarks are the same as in the SM;
due to this we can test the new physics beyond the SM predicted by this
particular model.

%%%%%%%%%%%%%%%%%%%%%%%%%%%%%%%%%%%%%%%%%%%%%%%%%%%%%%%%%%%%%%%%%%%%%%%%
\section {\label{sec:sec3}Fermion Masses and Mixing}
The Higgs scalars introduced in Sec.~\ref{sec:sec23} break the symmetry in 
a proper way and, at the same time, produce mass terms for the fermion 
fields via Yukawa interactions. 

In order to restrict the number of Yukawa couplings, and
produce an appealing mass spectrum, we introduce an anomaly-free discrete $Z_2$
symmetry~\cite{ross} with the following assignments of charges:
\begin{eqnarray}\label{z2} \nonumber
Z_2(Q^a_L,\phi_2,\phi_3,\phi_4,u^{ic}_L,d^{ac}_L, E^+_{lL})&=&1\\ 
Z_2(\phi_1, u^{3c}_L,U^{ic}_L,D^{c}_L, L_{lL}, l_{L}^+)&=&0,
\end{eqnarray}
where $a=1,2,3, \; i=1,2$ and $l=e,\mu,\tau$ are family indices as above.

Before entering into details let us mention that in some cases we may use a negative mass entry or find a negative mass eigenvalue, which are not troublesome, because we can always exchange the sign of the quark mass either by a change  of phase, or either by a transformation $\psi\longrightarrow\gamma_5\psi$ in the Weyl spinor $\psi$.

\subsection{\label{sec:sec31}The up quark sector}
The most general invariant Yukawa Lagrangean for the Up quark sector, without using the $Z_2$ symmetry, is given by 
\begin{eqnarray}\label{mup}\nonumber 
{\cal L}^u_Y&=&\sum_{i=1}^2[\sum_{\alpha =1,2,4}Q^i_L\phi_\alpha 
C(\sum_ah^{u\alpha}_{ia}u_L^{ac}+\sum_{j=1}^2 
h^{U\alpha}_{ij}U^{jc}_L)] \\  
& & + Q_L^3\phi_3^*C(\sum_{i=1}^2h^U_iU^{ic}_L +\sum_{a=1}^3h_a^uu_L^{ac}) + 
h.c., 
\end{eqnarray}
where the $h's$ are Yukawa couplings and $C$ is the charge conjugation 
operator. Then, in the basis $(u^1,u^2,u^3,U^1,U^2)$, and with the $Z_2$ symmetry enforced, 
we get from Eqs.(\ref{z2},\ref{mup}), the following tree-level Up quark
mass matrix:  
\begin{equation}\label{maup} M_u=\left(\begin{array}{ccccc} 
0 & 0 & h^{u2}_{13}v_2 & h_{11}^{U2}v_2 & h_{12}^{U2}v_2 \\ 
0 & 0 & h^{u2}_{23}v_2 & h^{U2}_{21}v_2 & h^{U2}_{22}v_2 \\ 
0 & 0 & h_3^{u}v_3 & h^{U}_1v_3 & h^U_2v_3\\ 
h_{11}^{u1}v_1 & h^{u1}_{12}v_1 & h^{u4}_{13}V & h_{11}^{U4}V & h_{12}^{U4}V\\ 
h_{21}^{u1}v_1 & h^{u1}_{22}v_1 & h^{u4}_{23}V & h_{21}^{U4}V & h^{U4}_{22}V \\ 
\end{array}\right), 
\end{equation} 
which is a see-saw type mass matrix. As a matter of fact, the analysis 
shows that the matrix $M_uM_u^\dagger$, for all the Yukawa coupling constants of order one but different 
from each other, and $v_1\approx v_2\approx v_3 <<V$, has 
the following set of eigenvalues: two of order $V^2$ associated with the two heavy 
exotic Up quarks, one of order $v_3^2$ associated with the top 
quark, and two see-saw eigenvalues of order $(v_iv_j/V)^2$ for $i,j=1,2,3$, related somehow to 
the quarks $u$ and $c$ in the first two families.

Also notice from matrix (\ref{maup}) that the permutation symmetry 
$u^1\leftrightarrow u^2$ imposed on the quarks of the first two 
families (which implies among other things that $h_{11}^{u1}=h_{12}^{u1}\equiv h_1^u$ and $h_{21}^{u1}=h_{22}^{u1}\equiv h_2^u$) conduces to a rank four see-saw 
type mass matrix, with the zero eigenvalue associated to the eigenstate 
$(1, -1, 0,0,0)/\sqrt{2}$. The $u^1\leftrightarrow u^2$ 
symmetry is thus related, in the context of this model, with a massless up type quark, that we identify with the $u$ quark in the first family.

In what follows, and without loss of generality, we are going to impose 
the condition $v_1=v_2=v_3\equiv v<<V$, with the value for $v$ fixed by 
the mass of the charged weak gauge boson 
$M_{W^\pm}^2=g_3(v_2^2+v_3^2)/2=g_3v^2$ which implies $v\approx 123$ GeV. Also, and in order to avoid proliferation of unnecessary parameters at this 
stage of the analysis, we propose to start with the following simple 
mass matrix 

\begin{equation}\label{maupp}
M_u^\prime=h_cv\left(\begin{array}{ccccc}
0 & 0 & 1 & 1 & 1 \\
0 & 0 & 1 & 1 & 1 \\
0 & 0 & h/h_c & 1 & 1 \\
1 & 1 & \delta^{-1} & h\delta^{-1}/h_c & \delta^{-1} \\
1 & 1 & \delta^{-1} & \delta^{-1} & \delta^{-1} 
\end{array}\right), 
\end{equation}
where $\delta = v/V$ is a perturbation expansion parameter and all the 
Yukawa coupling constants have been set equal to a common value $h_c$, 
except $h^u_3\equiv h$ which controls the top quark mass and $h_{11}^{U4}= h$ which simplifies the analysis and avoids democracy in the heavy sector.

The eigenvalues of $M_u^\prime M_u^{\prime\dagger}$, neglecting terms of order $\delta^3$ and higher are: a zero eigenvalue associated to the eigenvector $(1,-1,0,0,0)/\sqrt{2}$ that we identify with the up quark $u$ in the first family, a see-saw eigenvalue $4h_c^2v^2\delta^2$ related to the eigenvector 
\[[0,\eta, 0, -(h-h_c)^2\delta, (h-h_c)^2\delta]/N +{\cal O}(\delta^2),\]
where $\eta=1+2\delta (h+h_c)^2/(h-h_c)^2$ and $N$ is a normalization factor, both values associated with the charm quark $c$ in the second family; a tree-level value $(h-h_c)^2v^2/2 +{\cal O}(\delta^2)$ related to the eigenvector $[0,0,2\sqrt{2},-(h+h_c)\delta ,(h+3h_c)\delta]/N^\prime$ that we identify with the top quark $t$ in the third family. There are also two heavy values $(h-h_c)^2V^2$ and $(2h_ch+4h^2)V^2 + {\cal O}(\delta^2)$ associated with the two heavy states.

Using for for the top quark mass $m_t\approx 175$ GeV~\cite{pdb} we get $(h-h_c)\approx 2$, and using for the charm quark mass $m_c\approx 1.25$ GeV~\cite{pdb} we set the following bounds for the 3-3-1 mass scale:
2.5 TeV $\leq V\leq $100 TeV, for $0.1\leq h_c\leq 4$, a Yukawa coupling constant in the perturbative regime.

\begin{figure}
\includegraphics{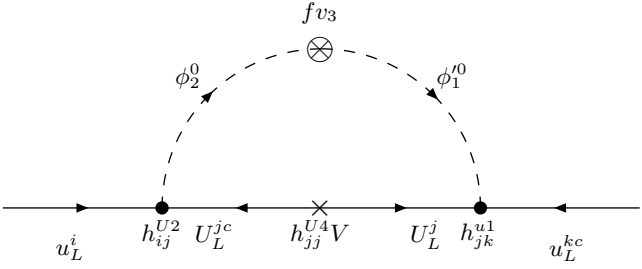}
\caption{\label{fig1}One loop diagram contributing to the radiative 
generation of the up quark mass. $i,j,k=1,2$ are indexes for the first two families.}
\end{figure}

The consistency of the model requires to find a mechanism able to produce a mass for the up quark $u$ in the first family, mass which is protected by the symmetry $u^1\leftrightarrow u^2$ between the quarks of the first two families. For this purpose the radiative mechanism~\cite{rad} can be implemented by using the rich scalar sector of the model. As a matter of fact, the two radiative diagrams depicted in Fig.~\ref{fig1} (one for $U^1$ and another for $U^2$) can be extracted from the Lagrangean, where the mixing in the Higgs sector in the diagram comes from a term in the scalar potential of the form $f\phi_1\phi_2\phi_3$.

The contribution from the two diagrams in Fig.~\ref{fig1} is finite and it is

\begin{eqnarray}\label{radup}
(M_u)_{ik}&=&fv_3N_{ik}[M^2m_1^2\ln (M^2/m_1^2)-\\ \nonumber
          & &M^2m_2^2\ln (M^2/m_2^2) + m_1^2m_2^2\ln (m_1^2/m_2^2)],
\end{eqnarray}
where 
\[N_{ik}=\frac{M\sum_j h_{ij}^{U2}h_{jj}^{U4}h_{jk}^{u1}}{[16\pi^2(m_2^2-m_1^2)(M^2-m_1^2)(M^2-m_2^2)},\]
with $M\approx V$ the mass of the exotic Up quark $U^j$ in the diagram, and $m_1$ and $m_2$ are the masses of $\phi_1^{\prime 0}$ and $\phi_2^0$ respectively. To estimate the contribution given by this diagram we assume the validity of the "extended survival hypothesis" (ESH)~\cite{esh} which in our case means $m_1\approx m_2 \approx v <<V$, producing a value
\[(M_u)_{ik}\approx -\frac{f\delta\ln\delta}{8\pi^2}\sum_{j=1}^2h_{ij}^{U2}h_{jj}^{U4}h_{jk}^{u1},\]
which for the symmetry $u^1\leftrightarrow u^2$ mentioned above implies a democratic type mass submatrix in the upper left $2\times 2$ mass matrix $M_u$. So, in order to produce a mass different from zero for the up quark in the first family, this symmetry must be slightly broken. The simplest way found to accomplish this breaking is to set $h_{21}^{u1}=1+k_u$ and $h_{12}^{U2}=1-k_u$, with $k_u$ a small parameter, and all the other Yukawa coupling constants as in matrix (\ref{maupp}) (this is what we mean by ``slightly broken"), with $k_u$ related to the $u$ quark mass $m_u$ which thus becomes 
\[m_u\approx -[\frac{(h+1)}{2}]^2k^2_uf\delta \ln\delta/(8\pi^2),\]
a positive value $(\ln\delta<0)$, which for $h\approx 1, \; V\approx 25$ TeV, and $v\simeq 123$ GeV implies $m_u\approx 0.3k^2_uf10^{-3}$. So, a value of $f\approx v$ (as implied by the ESH~\cite{esh})and $k_u=0.2$ implies $m_u\approx 1.5$ MeV, without introducing a new mass scale, neither a hierarchy in the Yukawa coupling constants of the Up quark sector of this particular model.

\subsection{\label{sec:sec32}The down quark sector}
The most general Yukawa terms for the Down quark sector, using the four 
Higgs scalars introduced in Sec.(\ref{sec:sec22}), are 
\begin{eqnarray}\label{mdown} \nonumber 
{\cal L}^d_Y&=& \sum_{i=1}^2Q^i_L\phi_3C(\sum_{a=1}^3h_{ia}^dd_L^{ac} + 
h_i^{\prime D}D_L^c) \\
&+&Q_L^3\sum_{\alpha=1,2,4}\phi_\alpha^*(h^D_\alpha D^c_L + 
\sum_{a=1}^3h_{a\alpha}^dd_L^{ac}) + h.c.. 
\end{eqnarray}
In the basis $(d^1,d^2,d^3,D)$ and using the discrete symmetry $Z_2$, the 
former expression produces a $4\times 4$ mass matrix with two zero eigenvalues, one see-saw eigenvalue associated  with the bottom quark $b$ in the third family, and a heavy eigenvalue  of order $V$ related with the exotic quark $D$. Unfortunately, the $Z_2$ symmetry used, allows to the right-handed ordinary Down quarks $d^{ac}_L$ to couple only to $\phi_1$ in a vertex where only $Q^3_L$ is present; as a consequence, any set of radiative diagrams able to provide mass terms to the Down quarks in the first two families, ends up in democratic type mass submatrices in the $(d^1,d^2,d^3)$ subspace, and the rank of the mass matrix can not be changed.

The simplest way found to provide with masses for the down $d$ and strange $s$ quarks in the context of this model, is to add new ingredients. We proposse to add first an extra Down exotic quark $D^\prime$, with quantum numbers $D^\prime_L\sim (3,1,-1/3), \; D^{\prime c}_L\sim (3^*,1,1/3)$ (which by the way do not affect the anomaly cancellation in the model because it belongs to a vectorlike representation). Also, and in order to implement the see-saw mechanism for this new exotic quark, we introduce a neutral scalar field $\phi_5^0\sim (1,1,0)$ with VEV $<\phi^0_5>=v_5\approx v$ (which does not contribute to the $W^\pm$ mass). The $Z_2$ charges of the new fields are all zero.

With the new fields, and in the basis $(d^1,d^2,d^3,D,D^\prime)$, the following $5\times 5$ mass matrix is obtained:
\begin{equation}\label{madown}
M_d=\left(\begin{array}{ccccc}
0 & 0 & 0 & h_{1}^{\prime D}v_3 &h_{13}^{D^\prime}v_3\\
0 & 0 & 0 & h_{2}^{\prime D}v_3 &h_{23}^{D^\prime}v_3 \\
0 & 0 & 0 & h_{2}^{D}v_2 &h_{32}^{D^\prime}v_2\\
h_{11}^dv_1 & h_{21}^dv_1 & h_{31}^dv_1 & h_4^DV &h_4^{D^\prime}V\\
h_5^1 v_5 & h_5^2v_5 & h_5^3v_5 & h^\prime M^\prime & hM \\
\end{array}\right),
\end{equation}
where $M^\prime \approx M$ are  bare masses introduced by hand, that we set of the order of $V$.

The matrix $M_d$ is again a see-saw type mass matrix, with the product $M_d M_d^\dagger$ having rank one. As the algebra shows, for the particular case $v_2=v_3$, the eigenvector related to the zero eigenvalue is proportional to 
\[[( h_{2}^{^\prime D}h_{32}^{D^\prime}-h_{23}^{D^\prime}h_2^D) , (h_{13}^{D^\prime}h_2^{D} 
- h_1^{^\prime D}h_{32}^{D^\prime}),(h_{1}^{^\prime D}h_{23}^{D^\prime}-h_{13}^{D^\prime}h_2^{\prime D})].\]

In what follows and in order to simplify matters we are going to set again 
$v_1=v_2=v_3=v_5\equiv v$, start with all the Yukawa coupling constants equal to a common value
$h_b$ and, in order to avoid democracy in the heavy sector, we are going to assume conservation of the heavy flavor in the $(D,D^\prime)$ basis, which means $h_4^{D^\prime}=h^\prime=0$. With this assumptions we get an hermitian Down quark mass matrix with two zero eigenvalues related to the eigenvectors $(1,-1,0,0,0)/\sqrt{2}$ and $(1,1,-2,0,0)/\sqrt{6}$, that we identify with the down $d$ and strange $s$ quarks of the first two families. There is also for such a matrix a see-saw eigenvalue $6h_bv\delta$ associated with the eigenvector $(1,1,1,-3\delta,-3\delta)/\sqrt{3+18\delta^2}$ that we identify with the bottom quark $b$ in the third family. The other two eigenvalues of the matrix are of order $V$.

Notice from this analysis that $m_b/m_c\approx 3h_b/h_c$ without a hierarchy between $h_b$ and $h_c$.

The matrix $M_d$, with the constraints discussed in the previous section, can not either generate radiative masses for the quarks in the first two families, due to the flavor symmetry $d^1\leftrightarrow d^2\leftrightarrow d^3$ present. To generate masses for them such a symmetry must be broken. Working in this direction, let us partially break the symmetry, keeping at this stage the $d^1\leftrightarrow d^2$ symmetry. This is achieved by putting all the Yukawa coupling constants equal to a common value $h_b$, except $h_{32}^{D^\prime}=h_5^3\equiv h_s=h_d(1+k_s)$, where $k_s$ is a number smaller than one, related to the strange quark mass (when $k_s=0$, $m_s=0$).

The new orthogonal mass matrix generated in this way is a see-saw rank four mass matrix, with the zero eigenvalue related to the eigenvector $(1,-1,0,0,0)/\sqrt{2}$ associated with the down quark $d$ in the first family. The two see-saw eigenvalues are 
\[h_b v\delta (6 + 2k_s+k_s^2\pm \sqrt{36+24k_s+8k_s^2+4k_s^3+k_s^4})/2,\]
producing $m_b\approx h_b v\delta (6+2k_s+k_s^2/3)$ and $m_s\approx 2h_b v\delta k_s^2/3$, which implies $k_s\approx 3\sqrt{m_s/m_b} \approx 0.47\approx (h_s/h_b-1)$, where $m_s\approx 120$ GeV and $m_b\approx 4.8$ GeV were used~\cite{pdb}. From the former analysis we get $h_s\approx 1.47 h_b$ without a hierarchy between $h_s$ and $h_b$.

\begin{figure}
\includegraphics{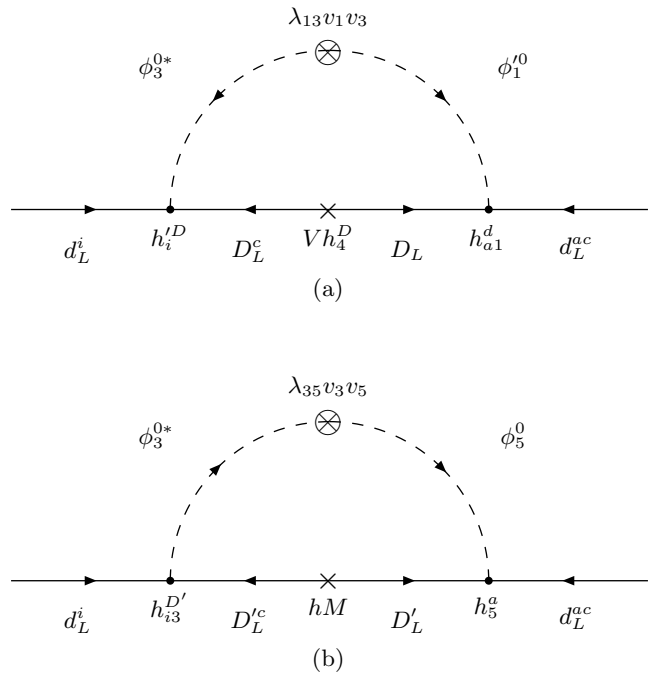}
\caption{\label{fig2}One loop diagram contributing to the radiative 
generation of the down quark mass. As in the main text, $i=1,2$ and $a=1,2,3$ are family indexes.}
\end{figure}

Finally, radiative diagrams able to produce nonzero mass for the quark $d$ in the first family, must be found. For this purpose the two diagrams in Fig.~\ref{fig2} can be extracted from the most general Lagrangean, where the scalar mixing are coming from terms in the scalar potential of the form $\lambda_{13}(\phi_1\phi_1^*)(\phi_3\phi_3^*)$ for the upper diagram and $\lambda_{35}(\phi_3\phi_3^*)(\phi_5\phi_5^*)$ for the lower one (two more diagrams using the  $u^3_L$ mass entries $h_3^uv_3$ in the fermion propagator are of the same order of the two diagrams depicted in Fig~\ref{fig2}, because the charged Higgs scalars mixing are proportional to $\lambda v_1V$).

In order to avoid hierarchies in the coupling constants $\lambda_{13}\approx\lambda_{35}\approx 1$ is going to be used. Again, democracy in the first two families is avoided by breaking the $d^1\leftrightarrow d^2$ symmetry which is achieved by using 
$h_1^{\prime D}\approx 1-k_d$ and $h_{11}^{d}\approx 1+k_d$, with $k_d$ a small number of order $10^{-1}$, related to the $d$ quark mass by the relation:
\[m_d\approx -h_b^2k_d^2v\delta\ln(\delta)/4\pi^2\approx 2\frac{h_b^2k_d^2}{h_U^2k_u^2}m_u,\]
which for $m_u\approx 3$ MeV and $m_d\approx 6$ MeV~\cite{pdb} implies $k_d\approx \sqrt{1.5}k_u$.

\subsection{\label{sec:sec33}{The quark mixing matrix}}
For a model like the one studied here, the ordinary quark mixing matrix $V_{mix}$ becomes the upper left $3\times 3$ submatrix of the unitary $5\times 5$ matrix  $V=V_L^uV_L^{d\dagger}$, where $V_L^u$ and $V_L^d$ are unitary matrices that diagonalize $M_uM_u^\dagger$ and $M_dM_d^\dagger$  respectively. As a consequence, $V_{mix}$ fails to be unitary, and special attention must be paid to the constraints coming from the experimental results which imply minimal mixing for the known quarks.

From the experimental side, the known results show that the $3\times 3$ quark mixing matrix, parametrized as 
\begin{equation}\label{vmix}
V_{mix}=\left(\begin{array}{ccc}
V_{ud} & V_{us} & V_{ub} \\
V_{cd} & V_{cs} & V_{cb} \\
V_{td} & V_{ts} & V_{tb} \\
              \end{array}\right),
\end{equation}
is almost diagonal, with measured values given by~\cite{pdb}

\[\left(\begin{array}{ccc}
0.9728\pm 0.0030 & 0.2257\pm 0.0021  & (36.7\pm 4.7)\times 10^{-4} \\
0.230\pm 0.011 &0.957 \pm 0.095 & (41.6\pm 0.6)\times 10^{-3} \\
(1.0\pm 0.1)\times 10^{-2}& (41.0\pm 3.0)\times 10^{-3} & > 0.78 \\
              \end{array}\right)\]
(notice that we are quoting the most uncertain direct measured values and not the values constrained by the unitary of $V_{CKM}$, the Cabbibo-Kobayashi-Maskawa (CKM) mixing matrix~\cite{ckm}).

One more complication in the frame of this model comes from the fact that our mass matrices may be flavor democratic in the Down quark sector but not in the Up quark sector, due to the three non zero tree-level top quark mass entries present in (\ref{maup}). By fortune, the model is full of free parameters and this last inconvenience can be circumvented by letting $h_{11}^{U4}, h_{12}^{U4}, h_{21}^{U4}$ and $h_{22}^{U4}$ to become free parameters in the interval $0.1\leq |h_{ij}^{U4}|\leq 4, \; i,j=1,2$. The numerical analysis shows that $h_{12}^{U4}=h_{21}^{U4}=0$ instead of one, is a more appropriate set of values in order to reproduce the experimental values of $V_{mix}$; unfortunately, the analytical results are not quite so neat for this last set of values, as compared with the previous quoted results.

The analysis also shows that violation of unitary in this model is proportional to $\delta^2$ and so, a large 3-3-1 scale, should reproduce fairly well the measured experimental results. The numerical analysis shows that, for 
$h_{12}^{U4}=h_{21}^{U4}=0, \; h_{11}^{U4}=h_{22}^{U4}=1$ and $V\approx 100$ TeV, with the other parameters as in the two previous sections, reproduce not only the experimental quoted values for $V_{mix}$, but also all the unitary constrained values of the $V_{CKM}$ mixing matrix. Lowering down the 3-3-1 scale to 60 TeV, keeping all the other parameters as above, reproduces also all the experimental unitary constraints values of $V_{CKM}$, except $V_{td}$ which turns out to be $(1.3\pm 0.3)\times 10^{-2}$ (statistical error) instead of the unitary value of (8.14+0.32-0.64)$\times 10^{-3}$ quoted in Ref.~\cite{pdb}, which predicts a $\beta$ angle 1.2 larger in the 
$V_{ud}V^*_{ub}+V_{cd}V^*_{cb}+V_{td}V^*_{tb}=0$ unitary triangle, result reflected already indirectly in the large $B^0-\bar{B}^0$ mixing measured at the B-factories~\cite{pdb}. But such a large 3-3-1 scale is a price too high to be payed, and renders at the end with a model unable to be tested in the upcoming generation of accelerators.

What we proposse at this point is to perform a numerical analysis using $h_{12}^{U4}, \; h_{21}^{U4}, \;h_{11}^{U4},\; h_{22}^{U4},\; h_b, \; k_u,\; k_s$ and $k_d$ as aleatory variables, with all the other Yukawa coupling constants equal to one, except $h$ which fixes the top quark mass, $h_c$ which fixes the 3-3-1 mass scale and $h_b$ which fixes the bottom quark mass. The analysis is constrained by the six quark mass values and the experimental measured values of $V_{mix}$, but not by the values obtained by imposing unitary in the $V_{CKM}$ mixing matrix. Then we look for the predictions of the model.

The ramdon numerical analysis using Mathematica Monte Carlo subroutines showed that, at the 3-3-1 scale of 10 TeV, the following set of parameters 
$h_{12}^{U4}=h_{21}^{U4}=0.26, \;h_{11}^{U4}=h_{22}^{U4}=-0.96,\; k_u=-0.15,\; k_s=0.38$ and $k_d=0.17$, reproduces the values of the $V_{CKM}$ with unitary constraints, except for three of them: $V_{td}=(1.1 \pm 0.2)\times 10^{-2}$, $V_{ub}=(45.8\pm 5)\times 10^{-4}$ and $V_{cb}=(40.2\pm 0.8)\times 10^{-3}$ (all the errors are statistical), which implies a large $B^0-\bar{B}^0$ mixing coming from $V_{td}$ and a depletion of the branching decay  $b\rightarrow s\gamma$ coming from $V_{cb}$; decay described by the magnetic dipole transition which is proportional to\cite{ndhg} $M_{b\rightarrow s\gamma}\sim V_{cb}V_{cs}^*$, with a value of $(42.21+0.10-0.80)\times 10^{-3}$ quoted for $V_{cb}$ in Ref.~\cite{pdb}.

\subsection{\label{sec:sec34}The charged lepton sector}
The most general Yukawa terms for the charged lepton sector, without using the $Z_2$ symmetry, is
\begin{equation} \label{mlep}
{\cal L}_Y^l=\sum_{\alpha = 1,2,4}\sum_{l,l^\prime = e,\mu,\tau} 
L_{lL}\phi^*_\alpha C(h_{ll^\prime}^{E\alpha}E_{l^\prime}^+
+h_{ll^\prime}^{e\alpha}l^{\prime +})_L + h.c., 
\end{equation}
which in the basis $(e,\; \mu, \; \tau, \; E_e, \; E_\mu, \; E_\tau)$ 
and with the discrete symmetry in Eq.~(\ref{z2}) enforced, produces the 
following $6\times 6$ mass matrix 
\begin{equation}\label{malep}
M_e=\left(\begin{array}{cccccc}
0 & 0 & 0 & h^{E2}_{ee}v_2 & h^{E2}_{e\mu}v_2 & 
h^{E2}_{e\tau}v_2 \\
0 & 0 & 0 & h^{E2}_{\mu e}v_2 & h^{E2}_{\mu\mu}v_2 & 
h^{E2}_{\mu\tau}v_2 \\
0 & 0 & 0 & h^{E2}_{\tau e}v_2 & h^{E2}_{\tau\mu}v_2 & 
h^{E2}_{\tau\tau}v_2 \\
h^{e1}_{ee}v_1 & h^{e1}_{e\mu}v_1 & h^{e1}_{e\tau}v_1 & h^{E4}_{ee}V & h^{E4}_{3e\mu}V  & 
h^{E4}_{e\tau} \\
h^{e1}_{\mu e}v_1 & h^{e1}_{\mu\mu}v_1 & h^{e1}_{\mu\tau}v_1 & h^{E4}_{\mu e}V & 
h^{E4}_{\mu\mu}V & h^{E4}_{\mu\tau}V \\
h^{e1}_{\tau e}v_1 & h^{e1}_{\tau\mu}v_1 & h^{e1}_{\tau\tau}v_1 &h^{E4}_{\tau e}V & 
h^{E4}_{\tau\mu}V  & h^{E4}_{\tau\tau}V \\
\end{array}\right)
\end{equation}
where again $v_1=v_2=v_3=v<<V$ is going to be used.
Assuming for simplicity conservation of the family lepton number in the 
exotic sector ($h^{E4}_{ll^\prime}=h_l\delta_{ll^\prime}$ which does not affect 
at all the main results), the matrix (\ref{malep}) still remains with 21 
Yukawa coupling constants and it is full of physical possibilities. For 
example, if all the 21 Yukawa coupling constants are different to each 
other (but of order one), we have that $M_eM_e^\dagger$ is a rank zero 
mass matrix, with three eigenvalues of order $V^2$ and three see-saw 
eigenvalues of order $v^2\delta^2$.

To start the analysis let us imposes the symmetry $e\leftrightarrow\mu\leftrightarrow\tau$, make all the Yukawa coupling constants equal to a common value $h_\tau$ and use conservation of the family lepton number in the exotic sector. With these assumptions the following orthogonal mass matrix is obtained:

\begin{equation}\label{maalep}
M_e^\prime=h_\tau v\left(\begin{array}{cccccc}
0 & 0 & 0 & 1 & 1 & 1 \\
0 & 0 & 0 & 1 & 1 & 1 \\
0 & 0 & 0 & 1 & 1 & 1 \\
1 & 1 & 1 & \delta^{-1} & 0 & 0 \\
1 & 1 & 1 & 0 & \delta^{-1} & 0 \\
1 & 1 & 1 & 0 & 0 & \delta^{-1} \\
\end{array}\right), 
\end{equation}
which is a symmetric rank four see-saw mass matrix, with the six 
eigenvalues given by
\begin{equation}\label{maasee}
h_lv [0,0,(\delta^{-1}\pm\sqrt{36+\delta^{-2}})/2,V,V],
\end{equation}
with the two zero eigenvalues related to the null subspace 
$(1,-1,0,0,0,0)/\sqrt{2}$ and $(1,1,-2,0,0,0)/\sqrt{6}$ that we identify in first approximation with the electron and the muon states (resembling the Down quark sector). Equations (\ref{maalep}) and (\ref{maasee}) implies that the $\tau$ lepton may be identify approximately with the vector $(1,1,1,0,0,0)/\sqrt{3}$, up to mixing with the heavy exotic leptons [the exact eigenvector is $(1,1,1,-3\delta, -3\delta, -3\delta)/\sqrt{3+27\delta^2}]$, with a mass value $-m_\tau\approx 9h_\tau v\delta = 4.5h_\tau m_c/h_u$, which for $m_\tau\approx 1.777$ GeV implies the relationship $h_u\approx 3h_\tau$.

The next step is to break the $e\leftrightarrow\mu\leftrightarrow\tau$ symmetry but just in the $\tau$ sector, keeping for a while the $e\leftrightarrow\mu$ symmetry. This is simple done by letting $h^{E2}_{\tau\tau}=h^{e1}_{\tau\tau}\equiv h_\mu \neq 1$ but of order one, with all the other Yukawa coupling consants as in Eq.~(\ref{maalep}). We thus get a rank five orthogonal mass matrix, with two see-saw eigenvalues, and a zero mass eigenstate related to the eigenvector $(1,-1,0,0,0,0)/\sqrt{2}$ that we identify with the electron state. The two see-saw eigenvalues, neglecting terms of ${\cal O}(\delta^2)$, are given by
\begin{equation}
\frac{h_\tau v\delta}{2}[8+(\frac{h_\mu}{h_\tau})^2\pm(2+\frac{h_\mu}{h_\tau}\sqrt{12-4(h_\mu/h_\tau)+(h_\mu/h_\tau)^2}].
\end{equation}
Using for $m_\tau \approx 1777$Gev and $m_\mu\approx 107.7$ Mev \cite{pdb} we get $h_\mu \approx 2.87 h_\tau$, which in turn implies $m_\tau \approx 15.3 h_\tau v\delta\approx 7.6 h_\tau m_c/h_u$.

\begin{figure}
\includegraphics{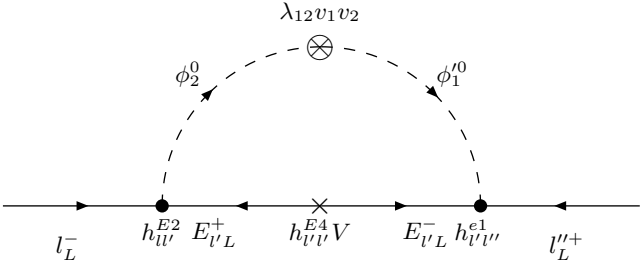}
\caption{\label{fig3}One loop diagram contributing to the radiative 
generation of the electron mass.}
\end{figure}

Again, radiative corrections able to generate masses to the electron must be found. For that purpose the three diagrams in Fig.~\ref{fig3} can be extracted from the Lagrangian (one for each exotic charged lepton), where the mixing in the Higgs sector is coming from a term in the scalar potential of the form $\lambda_{12}(\phi_1\phi^*_1)(\phi_2\phi_2^*)$. There are two more diagrams coming from the terms $f\phi_1\phi_2\phi_3$ and $f^\prime\phi_1\phi_3\phi_4$ which are proportional to $v_3(f^\prime -f)$ that can be  neglected under the assumption $f^\prime =f\approx v$.

The contribution given by this diagram, again under the assumption of validity of the ESH~\cite{esh} is 
\[(M_e^\prime)_{ll^\prime}\approx \frac{\lambda_{12}\delta\ln\delta}{8\pi^2}\sum_{l^\prime}h_{ll^\prime}^{E2}h_{ll^{\prime\prime}}^{e1},\]
that for the particular values of the Yukawa coupling constants in matrix (\ref{maalep}) generate a democratic mass submatrix in the $2\times 2$ upper left corner of $M_e^\prime$. Again, the alternative we have at hand is to softly break the $e\leftrightarrow\mu$ symmetry present in the mass matrix (\ref{maalep}). This is achieved by letting $h_{ee}^{E2}\approx 1-k_e$ and $h_{ee}^{e1}\approx 1+k_e$, with $k_e\sim 10^{-1}$ as before.

The evaluation of the diagram in Fig.~\ref{fig3} gives 
\[m_e\approx -\frac{\lambda_{12}v\delta\ln\delta k_e^2}{4\pi^2},\]
which for $m_e=0.51$ MeV~\cite{pdb}, $\lambda_{12}\approx 1$, $V\approx 25$ TeV, and $v=124$ GeV, produces a value of $k_e\approx 0.08$, in agreement with our original assumption.

\subsection{\label{sec:sec35}The neutral lepton sector} 
With the particle content introduced so far there are not tree-level mass terms for the neutrinos. Masses for the neutral lepton sector are obtained only by enlarging the model with extra fields, which may implement one or several of the following mechanisms:
\subsubsection{\label{sec:sec341}Tree-Level masses}
In the context of the model studied here, tree-level masses for neutrinos can be generated only by introducing scalar fields belonging to irrep $\{6^*\}$ of $SU(3)_L$. These scalars can be written as the $3\times 3$ symmetric tensor 
\begin{equation}
\chi_{\{\alpha,\beta\}}=\left(\begin{array}{ccc}
\chi_{11}^{-4/3+X} & \chi_{12}^{-1/3+X}    & \chi_{13}^{-1/3+X} \\
            & \chi_{22}^{2/3+X} & \chi_{23}^{2/3+X} \\
            &                & \chi_{33}^{2/3+X} \\
\end{array}\right) \sim (1,6^*,X),
\end{equation}
where the upper symbol stands for the electric charge. Clearly, a VEV of the form $\langle\chi_{11}(1,6^*,4/3)^0\rangle\sim\omega$
is able to produce the following Majorana mass terms: $h_{l^\prime l}^\nu\omega\nu_{l^\prime}^0\nu_{lL}^0$. If so, $h_{l^\prime l}^\nu$ must become very small numbers, or either $\omega$ must be a new very small mass scale 
in order to cope with the experimental constraints~\cite{valle}, 
implying for the model a hierarchy in the Yukawa coupling constants, or either the introduction of a new mass scale for the model.
\subsubsection{\label{sec:sec342}See-Saw masses}
The see-saw mechanism can be implemented in the model by adding a singlet, electrically neutral Weyl spinor $N^0_L\sim (1,1,0)$ with $Z_2$ charge 1, which picks up a tree-level mass value $V^\prime N_L^0N_L^0$ with $V^\prime$ an undetermined mass scale. Then, a Yukawa Lagrangian of the form \[\sum_lh_lL_{lL}\phi_3^*N_L^0,\] 
will produce a see-saw type mass matrix 
\begin{equation}\label{maneu}
M_\nu=\left(\begin{array}{cccc}
0 & 0 & 0 & h_ev_3 \\
0 & 0 & 0 & h_\mu v_3 \\
0 & 0 & 0 & h_\tau v_3 \\
h_ev_3 & h_\mu v_3 & h_\tau v_3 & V^\prime \\
\end{array}\right),
\end{equation}
which has two nonzero tree-level mass eigenvalue 
$V^\prime\pm\sqrt{V^{\prime 2} + 4(h_e^2+h_\mu^2+h_\tau^2)v_3^2}$, one of them of the see-saw type and proportional to $v_3^2/V^\prime$ which for a convenient larger value of $V^\prime$ (again a new mass scale) produces a small neutrino mass. Of course, this mechanism alone is not enough to explain the spectrum because two neutrinos will remain massless, something which is ruled out by experimental results~\cite{valle}.
\subsubsection{\label{sec:sec343}Radiative masses}
Radiative Majorana masses for the neutrinos are generated when a new scalar triplet 
$\phi_5=(\phi_5^{++}, \phi_5^+ ,\phi_5^{\prime +})\sim (1,3,4/3)$ is introduced, with a $Z_2$ charge 
equal to zero (notice that $\langle\phi_5\rangle\equiv 0$). This new scalar triplet couple to the spin 1/2 leptons via a term in the Lagrangian of the form:
\begin{eqnarray*}\nonumber
{\cal L} &=& \sum_{ll^\prime}h^\nu_{ll^\prime}L_{lL}L_{l^\prime L}\phi_5 = 
\sum_{ll^\prime}h^\nu_{ll^\prime}[\phi_5^{++}(l^-_LE^-_{l^\prime L}-l^{\prime -}_LE^-_{lL}) \\ \nonumber 
&+& \phi_5^{+}(E^-_{lL}\nu_{l^\prime L}-E^-_{l^\prime L}\nu_{lL})
+\phi_5^{\prime +}(\nu_{lL}l^{\prime -}_L-\nu_{l^{\prime}L}l^-_L)],
\end{eqnarray*}
for $l\neq l^\prime = e,\mu,\tau$.

\begin{figure}
\includegraphics{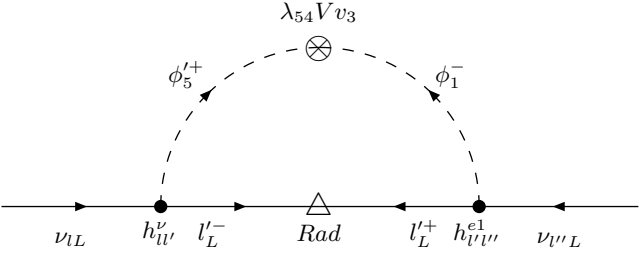}
\caption{\label{fig4}Loop diagrams contributing to the radiative 
generation of Majorana masses for the neutrinos.}
\end{figure}

Using $\phi_5$, the following terms in the scalar potential Lagrangian are allowed by the $Z_2$ discrete symmetry
\[\lambda_{51}(\phi_5.\phi_1^*)(\phi_3.\phi_2^*); \; \lambda_{52}(\phi_5.\phi_1^*)(\phi_3.\phi_4^*);\]
\[\lambda_{53}(\phi_5.\phi_2^*)(\phi_3.\phi_1^*); \;  \lambda_{54}(\phi_5.\phi_4^*)(\phi_3.\phi_1^*).\]
The former expressions allow to draw the radiative diagram depicted in Fig.~\ref{fig4}, which is the only diagram available for the radiative mechanism in the neutral lepton sector. 

Notice by the way that this diagram is already a second order radiative diagram because its charged lepton mass insertion is already a first order radiative correction (see the diagram in Fig.~\ref{fig3}) and its value is smaller than the value produced by any other radiative diagram already studied in this paper. Attempts to draw a diagram with the exotic heavy leptons in the fermion propagator became fruitless, due to the $Z_2$ symmetry introduced in the analysis [a term like $(\phi_5.\phi_2^*)(\phi_3.\phi_4^*)$ is not $Z_2$ allowed!].

\subsubsection{\label{sec:sec344}The Zee-Babu mechanism}
Introducing a new $SU(3)_L$ singlet, electrically charged scalar, as it is done for example in Ref.~\cite{kita1}, the two loop diagram of the Zee-Babu mechanism~\cite{zeeba} can be included in the context of this model.

Without going into further details, let us say that the neutrino mass spectrum is outside the scope of the present analysis.

%%%%%%%%%%%%%%%%%%%%%%%%%%%%%%%%%%%%%%%%%%%%%%%%%%%%%

\section{\label{sec:sec4}Gauge Coupling Unification}
In a field theory, the coupling constants are defined as effective values, which are energy scale dependent according to the renormalization group equation. In the modified minimal substraction scheme \cite{mss}, which we adopt in what follows, the one loop renormalization group equation (RGE) for $\alpha = g^2/4\pi$ is given by

\begin{equation}\label{rge}
\mu\frac{d\;\alpha}{d\;\mu}\simeq -b\alpha^2,
\end{equation}

\noindent
where $\mu$ is the energy at which the coupling constant $\alpha$ is evaluated. The constant value $b$, called the beta function, is completely determined by the particle content of the model by 
\[2\pi b=\frac{11}{6}C(\mbox{vectors})-\frac{2}{6}C(\mbox{fermions})-\frac{1}{6}C(\mbox{scalars}),\]
where $C(\dots )$ is the group theoretical index of the representation inside the parentheses (we are assuming Weyl fermions and complex scalar fields ~\cite{slansky}).

For the energy interval $m_Z<\mu<M_G$, the one loop solutions to the RGE (\ref{rge}) for the three SM gauge coupling constants are
\begin{equation}\label{srge}
\alpha^{-1}_i(m_Z)=\frac{\alpha_i^{-1}(M_G)}{c_i}-b_i(F,H)\ln\left(\frac{M_G}{m_Z}\right), 
\end{equation}
where $i=Y,2,c$ refers to the coupling constants of $U(1)_Y,\; SU(2)_L$ and $SU(3)_c$ respectively, with the beta functions given by 
\begin{equation}
2\pi\left(\begin{array}{c} b_Y \\b_2 \\b_c \end{array}\right)=
\left(\begin{array}{c} 0 \\ \frac{22}{3} \\ 11 \end{array}\right)
-\left(\begin{array}{c} \frac{20}{9} \\ \frac{4}{3} \\ \frac{4}{3} \end{array}\right)F
-\left(\begin{array}{c} \frac{1}{6} \\ \frac{1}{6} \\ 0 \end{array}\right)H, 
\end{equation}
where $F$ is the number of families contributing to the beta functions and $H$ is the number of low energy $SU(2)_L$ scalar field doublets ($H=1$ for the SM). In Eq.~(\ref{srge}) the constants $c_i$ are group theoretical factors which depend upon the embedding of the SM factors into a covering group, and warrant the same normalization for the covering group $G$ and for the three group factors in the SM. For example, if the covering group is $SU(5)$, then $(c_Y,c_2,c_c)=(3/5,1,1)$, but they are different for other covering groups (see for example the Table in Ref.~\cite{lpz}).

The three running coupling constants $\alpha_i$ in Eq.~(\ref{srge}), may or may not converge into a single energy GUT scale $M_G$; if they do, then $\alpha_i(M_G)=\alpha$ is a constant independent of the index $i$. Now, for a given embedding into a fixed covering group, the $c_i$ values are fix, and if we use for $F=3$ (an experimental fact) and $H=1$ as in the SM, then  Eqs. (\ref{srge}) constitute a set of three equations with two unknowns, $\alpha$ and $M_G$ which may or may not have a consistent solution (more equations than unknowns). 

The inputs to be used in Eq. (\ref{srge}) for $\alpha^{-1}_i(m_Z)$ are calculated from the experimental results \cite{pdb} 
\begin{eqnarray}\nonumber
\alpha_{em}^{-1}(m_Z)&=& \alpha_Y^{-1}(m_Z)+\alpha_2^{-1}(m_Z)\\ \nonumber 
&=& 127.918\pm 0.018\\ \nonumber
\sin^2\theta_W(m_Z)&=&1-\alpha_Y^{-1}(m_Z)\alpha_{em}(m_Z) \\ \nonumber 
&=&0.23120\pm 0.00015\\ \nonumber
\alpha_c(m_Z)&=& 0.1213\pm 0.0018,
\end{eqnarray}
which imply $\alpha^{-1}_Y(m_Z)=98.343\pm 0.036, \;\alpha_2^{-1}(m_Z)=29.575 \pm 0.054$, and $\alpha_c^{-1}(m_Z)= 8.244 \pm 0.122$.

It is a well known fact that the model based on the non supersymmetric
$SU(5)$ group of Georgi and Glashow \cite{gg} lacks of gauge coupling
unification because $M_G$ calculated from the RGE is not unique in the range $10^{14}$ GeV $\leq
M_G \leq 10^{16}$ GeV, predicting for the proton lifetime $\tau_p$ a value
between $2.5\times 10^{28}$ years and $1.6\times 10^{30}$ years, values that are 
ruled out by experimental measurements~\cite{amaldi}. If we
introduce one more free parameter in the solutions to the RGE as for
example letting $H$ to become a free integer number, then we have now
three unknowns with three equations that always have mathematical solution
(not necessarily with physical meaning). Doing that in Eqs. (\ref{srge})
we find that for $H=7$ (seven Higgs doublets) we get the unique solution
$M_G=10^{13}$ GeV $>>m_Z$ which, altough a physical solution, it is ruled
out by the proton lifetime. So, if we still want unification, new physics
at an intermediate mass scale $M_V$ such that $m_Z<M_V<M_G$ must exists,
being supersymmetry (SUSY) a popular candidate for that purpose
~\cite{amaldi}.

The question now is if the 3-3-1 model under consideration in this paper,
introduces an intermediate mass scale $M_V$ such that it achieves proper
gauge coupling unification, being an alternative for SUSY. To answer this
question using $SU(6)$ as the covering group as presented in Section
\ref{sec:sec24}, we must solve the following set of seven equations:
\begin{eqnarray}\label{srge331}\nonumber
\alpha^{-1}_i(m_Z)&=&\frac{\alpha_i^{-1}(M_V)}{c_i}-b_i(F,H)\ln\left(\frac{M_V}{m_Z}\right)\\ \nonumber
\alpha^{-1}_j(M_V)&=&\frac{\alpha^{-1}}{c_j^\prime}-b_j^\prime\ln\left(\frac{M_G}{M_V}\right)\\ 
\alpha_Y^{-1}(M_V)&=&\alpha_1^{-1}(M_V)+\alpha_3^{-1}(M_V)/3,
\end{eqnarray}
where the last equation is just the matching conditions in Eq.(\ref{mc}),
and $i=Y,2,c$ and $j=1,3,c$ for the SM and the 3-3-1 model, respectively.
The constants $c_i$ are $(c_Y,c_2,c_3)=(3/5,1,1)$ as before, and
$(c_1^\prime,c_3^\prime,c_c^\prime)=(3/4,1,1)$, with the value $c_1^\prime
=3/4$ calculated from the electroweak mixing angle in Eq. (\ref{ewk}).
$b_j^\prime$ stand for the beta functions for the 3-3-1 model under study
here.

Eqs.~(\ref{srge331}) constitute a set of seven equations with seven unknowns $\alpha, \; \alpha_j(M_V), \; M_V, \; M_G$ and $\alpha_Y(M_V) \;\; [\alpha_2(M_V)=\alpha_3(M_V)$ according to the matching conditions]. There is always mathematical solution to this set of equations, but we want only physical solutions, that is solutions such that $m_Z<M_V<M_G$.

The new beta functions calculated with the particle content introduced in Sections \ref{sec:sec21}, \ref{sec:sec22}, \ref{sec:sec23} and \ref{sec:sec32} (it includes the new exotic Down quark $D^\prime$) are: 
\begin{equation}\label{beta331}
2\pi\left(\begin{array}{c} b_1^\prime \\b_3^\prime \\b_c^\prime \end{array}\right)=
\left(\begin{array}{c} 0-12-11/9 \\ 11-4-4/6 \\ 11-20/3-0 \end{array}\right)
=\left(\begin{array}{c} -119/9 \\ 19/3 \\ 13/3 \end{array}\right),
\end{equation}
where in the middle term we have separated the contributions coming from the gauge bosons, the fermion fields and the scalar fields in that order. When we introduce these values in Eq. (\ref{srge331}) we do not obtain a physical solution in the sense that we get $m_Z<M_G<M_V$.

Of course, if there are more particles at the 3-3-1 mass scale then the beta functions given in Eqs. (\ref{beta331}) are not the full story. In particular we know from Sec. \ref{sec:sec34} that at least new Higgs scalars are needed in order to generate a consistent lepton mass spectrum, so let us allow the presence in our model of the following Higgs scalar multiplets at the 3-3-1 mass scale: $N_X^{(1)} \; SU(3)_L$ singlets (with $U(1)_X$ hypercharge equal to $X$), $N_X^{(3)}$ triplets (color singlets), $\tilde{N}_X^{(3)}$ leptoquark triplets (color triplets) and $N_X^{(6)}$ sextuplets (color singlets). These new particles contribute to the beta functions $b^\prime_j$ with the extra values:

%\begin{twocolumn}
\begin{equation}\label{betap331}
2\pi\left(\begin{array}{c} b_1^\prime \\ b_3^\prime \\b_c^\prime \end{array}\right)=
-\left(\begin{array}{c} 
\sum_XX^2 f(N_X^{(6)}, \tilde{N}_X^{(3)}, N_X^{(3)},N_X^{(0)})\\  
{1 \over 6}\sum_X(N_X^{(3)}+3\tilde{N}_X^{(3)}+5N_X^{(6)}) \\ 
\sum_X\tilde{N}_X^{(3)}/2 \end{array}\right), 
\end{equation}
%\end{twocolumn}
where the function $f(\dots )$ is $f(N_X^{(6)}, \tilde{N}_X^{(3)}, N_X^{(3)},N_X^{(0)})=  
(2N_X^{(6)}+3\tilde{N}_X^{(3)}+N_X^{(3)}+N_X^{(0)}/3)$; 
with this new $SU(3)_L$ multiplets contributing or not to the beta functions $b_i$ of the SM factor groups, in agreement with the extended survival hypothesis \cite{esh} (for example, a sextuplet with a VEV $\langle\chi_{11}(1,6^*,4/3)\rangle \sim \omega$ contributes as an $SU(2)_L$ doublet in $b_Y$ and $b_2$, etc.).

\begin{figure*}
\includegraphics{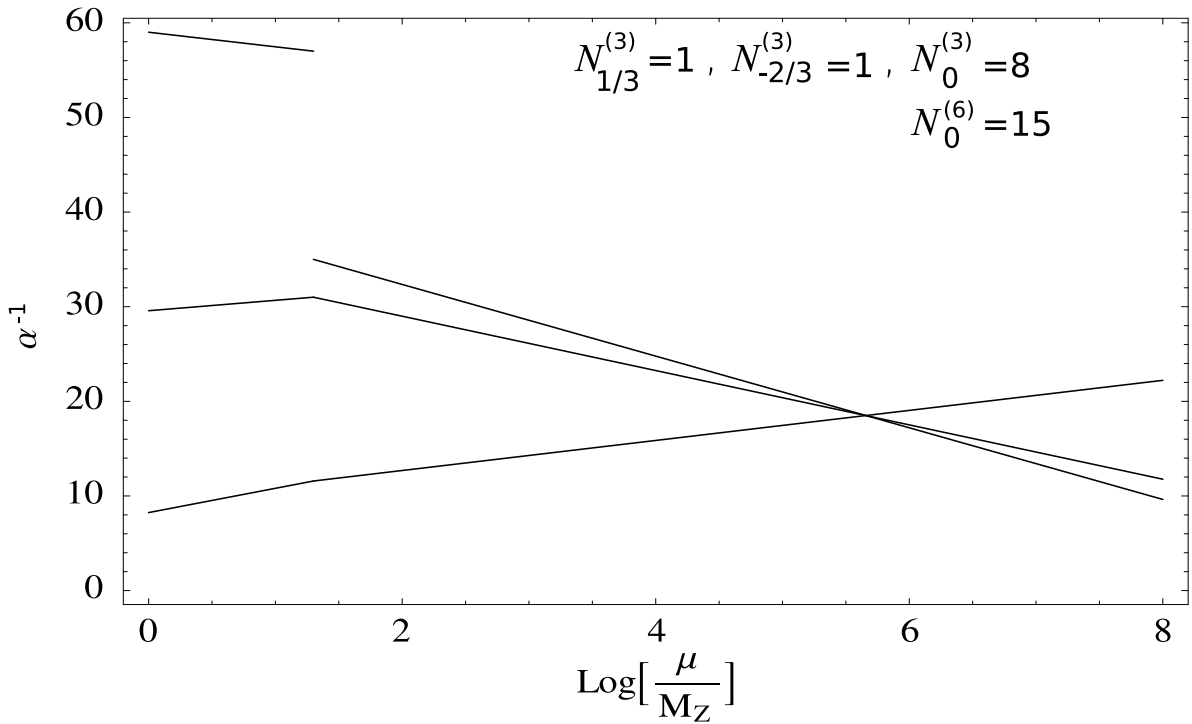}
\caption{\label{fig5}Solutions to the RGE for the 3-3-1 model. For the meaning of $N^{(r)}_X$ see the main text.}
\end{figure*}

The calculation shows that for the following set of extra scalar fields which do not develop VEV:  $N_X^{(0)}=0$, $N^{(3)}_{1/3}=1$, $N^{(3)}_{-2/3}=1$, $\tilde{N}^{(3)}_X=0$ $N^{(3)}_0=8$ and $N_0^{(6)}=15$, the set of equations in (\ref{betap331}) has the physical solution 
\begin{equation}\label{scales}
M_V\approx 2.0 \mbox{TeV}  < M_G\approx 3.0\times 10^{7} \mbox{GeV}, 
\end{equation}
which provides with a convenient 3-3-1 mass scale, and a low unification GUT mass scale, as it is shown in Fig.~\ref{fig3}.

But, is this low GUT scale in conflict with proton decay? The answer is not, because due to the $Z_2$ symmetry our unifying group is $SU(6)\times Z_2$. Then we must assign to each irrep of $SU(6)$ in Eq.~(\ref{reducible}) a given $Z_2$ value in accord with the $Z_2$ value assigned to the 3-3-1 states in Eq.~(\ref{z2}). For example, if we assign to one of the four $\{6^*\}=\{D^c, -N_E^0,E^-,N_E^{0c}\}_L$ states in (\ref{reducible}) a $Z_2$ value equal to 1, then we can perfectly identify $D_L^c$ with one of the ordinary down quarks $(d^c,s^c,b^c)_L$, but then $(-N_E^0, E^-, N_E^{0c})_L$ can not correspond to $(-\nu_l^0, l^-,\nu_l^{0c})_L$ because all of them have a $Z_2$ value equal to zero; and the same for the other way around. As a consequence, the down quark $d_L^c$ can not live together with $(\nu_e,e^-)_L$ in the same $SU(6)\times Z_2$ irrep, and the proton can not decay to light states belonging to the weak basis. The decay can of course occur via the mixing of ordinary 3-3-1 states with the extra new states in $SU(6)$, but such a mixing is of the order of $(M_V/M_G)^2$ which is a very small value. Of course, this argument is valid as far as we can find a mechanism able to produce GUT scale masses for all the extra states, but such analysis is outside the present work.

%%%%%%%%%%%%%%%%%%%%%%%%%%%%%%%%%%%%%%%%%%%%%%%%%%%%%%%%%%%%%%%%
\section{\label{sec:sec5}Constraints on the parameters}
In this section we are going to set bounds on the mass of the new neutral gauge boson $Z_2^\mu$, and its mixing angle with the ordinary neutral gauge boson $Z_1^\mu$. We also are going to set constraints coming from unitary violation of the quark mixing matrix and the possible existence of FCNC effects.

\subsection{\label{sec:sec51}Bounds on $M_{Z_2}$ and $\theta$.}
The diagonalizing of the quark mass matrices presented in sections (\ref{sec:sec31}) and (\ref{sec:sec32}) allow us to identify the mass eigenstates as a function of the flavor states. This information is going to be used next, in order to set proper bounds for $\sin\theta$, the mixing angle between the two neutral currents, and $M_{Z_2}$, the mass of the new neutral gauge boson. In the analysis we are going to include the $c$ and $b$ quark couplings to $Z_1^\mu$, values measured with good accuracy at the $Z$ pole from CERN $e^+e^-$ collider (LEP)~\cite{pdb}. Experimental measurements from the SLAC Linear Collider (SLC), and
atomic parity violation are also going to be taken into account. The set of experimental constraints used are presented in Table \ref{tab3}.

The expression for the partial decay width for $Z^{\mu}_1\rightarrow
f\bar{f}$, including only the electroweak and QCD virtual corrections is
\begin{eqnarray}\label{gamma}\nonumber
\Gamma(Z^{\mu}_1\rightarrow f\bar{f})&=&\frac{N_C G_F
M_{Z_1}^3}{6\pi\sqrt{2}}\rho \Big\{\frac{3\beta-\beta^3}{2}
[g(f)_{1V}]^2 \\ \label{ancho}
& + & \beta^3[g(f)_{1A}]^2 \Big\}(1+\delta_f)R_{EW}R_{QCD}, \quad
\end{eqnarray}
\noindent 
where $f$ is an ordinary SM fermion, $Z^\mu_1$ is the physical gauge boson
observed at LEP, $N_C=1$ for leptons while for quarks
$N_C=3(1+\alpha_s/\pi + 1.405\alpha_s^2/\pi^2 - 12.77\alpha_s^3/\pi^3)$,
where the 3 is due to color and the factor in parenthesis represents the
universal part of the QCD corrections for massless quarks 
(for fermion mass effects and further QCD corrections which are 
different for vector and axial-vector partial widths see 
Ref.~\cite{kuhn}); $R_{EW}$ are the electroweak corrections which include 
the leading order QED corrections given by $R_{QED}=1+3\alpha/(4\pi)$. 
$R_{QCD}$ are further QCD corrections (for a comprehensive review see 
Ref.~\cite{leike} and references therein), and $\beta=\sqrt{1-4 m_f^2/
M_{Z_1}^2}$ is a kinematic factor which can be taken equal to $1$ for all 
the SM fermions except for the bottom quark. 
The factor $\delta_f$ contains the one loop vertex
contribution which is negligible for all fermion fields except for the 
bottom quark for which the contribution coming from the top quark at the 
one loop vertex radiative correction is parametrized as $\delta_b\approx 
10^{-2} [1/5-m_t^2/(2 M_{Z_1}^2)]$ \cite{pich}. 

The $\rho$ parameter can be expanded as $\rho = 1+\delta\rho_0 + \delta\rho_V$ where the 
oblique correction $\delta\rho_0$ is given by
$\delta\rho_0\approx 3G_F m_t^2/(8\pi^2\sqrt{2})$, and $\delta\rho_V$ is 
the tree level contribution due to the $(Z_{\mu} - Z'_{\mu})$ mixing which 
can be parametrized as $\delta\rho_V\approx
(M_{Z_2}^2/M_{Z_1}^2-1)\sin^2\theta$. Finally, $g(f)_{1V}$ and $g(f)_{1A}$
are the coupling constants of the physical $Z_1^\mu$ field with ordinary
fermions which for this model are listed in Table \ref{tab1}.

Notice that in our expression for $\Gamma(Z_1^\mu\longrightarrow f\bar{f})$ in Eq.~(\ref{gamma}), the 3-3-1 contributions are kept at tree-level, which as a first approximation is correct due to the fact that $\delta\rho_0(331)\approx 0$, since only $SU(2)_L$ Higgs scalar singlets and doublets develop VEV~\cite{gluza}.

In what follows we are going to use the experimental values \cite{pdb}:
$M_{Z_1}=91.188$ GeV, $m_t=174.3$ GeV, $\alpha_s(m_Z)=0.1192$,
$\alpha(m_Z)^{-1}=127.938$, and $\sin\theta^2_W=0.2333$. The experimental
values are introduced using the definitions $R_\eta\equiv
\Gamma_Z(\eta\eta)/\Gamma_Z(hadrons)$ for $\eta=e,\mu,\tau,b,c,s,u,d$.

As a first result notice from Table \ref{tab1} that our model predicts 
$R_e=R_\mu=R_\tau$, in agreement with the experimental results in Table 
\ref{tab3}, independent of any flavor mixing at tree-level.

The effective weak charge in atomic parity violation, $Q_W$, can be 
expressed as a function of the number of protons $(Z)$ and the number of 
neutrons $(N)$ in the atomic nucleus in the form 

\begin{equation}
Q_W=-2\left[(2Z+N)c_{1u}+(Z+2N)c_{1d}\right], 
\end{equation}
\noindent
where $c_{1q}=2g(e)_{1A}g(q)_{1V}$. The theoretical value for $Q_W$ for 
the Cesium atom is given by \cite{sirlin} $Q_W(^{133}_{55}Cs)=-73.09\pm0.04 
+ \Delta Q_W$, where the contribution of new physics is included in $\Delta 
Q_W$ which can be written as \cite{durkin}

\begin{equation}\label{DQ} 
\Delta 
Q_W=\left[\left(1+4\frac{S^4_W}{1-2S^2_W}\right)Z-N\right]\delta\rho_V
+\Delta Q^\prime_W.
\end{equation}

The term $\Delta Q^\prime_W$ is model dependent and it can be obtained for
our model by using $g(e)_{iA}$ and $g(q)_{iV}$, $i=1,2$, from Tables 
\ref{tab1} and \ref{tab2}. The value we obtain is

\begin{equation}
-\Delta Q_W^\prime=(9.16 Z + 4.94 N) \sin\theta + (4.63 Z + 3.74 N)
\frac{M^2_{Z_1}}{M^2_{Z_2}}\; .
\end{equation}

The discrepancy between the SM and the experimental data for $\Delta Q_W$ 
is given by \cite{casal}

\begin{equation}
\Delta Q_W=Q^{exp}_W-Q^{SM}_W=0.45\pm 0.48,
\end{equation}
which is $1.1\; \sigma$ away from the SM predictions.
\begin{table}
\caption{\label{tab3}Experimental data and SM values for some parameters 
related with neutral currents.}
\begin{ruledtabular}
\begin{tabular}{lcl}
& Experimental results & SM \\ \hline
$\Gamma_Z$(GeV)  & $2.4952 \pm 0.0023$  &  $2.4966 \pm 0.0016$  \\   
$\Gamma(had)$ (GeV)  & $1.7444 \pm 0.0020$ & $1.7429 \pm 0.0015$ \\ 
$\Gamma(l^+l^-)$ (MeV) & $83.984\pm 0.086$ & $84.019 \pm 0.027$ \\
$\Gamma(inv)(MeV)$  & $499.0\pm 1.5$  & $501.81\pm 0.13$ \\
$\frac{\Gamma(b\rightarrow s\gamma)}{\Gamma(b\rightarrow Xe\nu)}$ & $3.39^{+0.62}_{-0.54}\times 10^{-3}$ & $(3.23\pm 0.09)\times 10^{-3}$ \\
$R_e$ & $20.804\pm 0.050$ & $20.744\pm 0.018$ \\ 
$R_\mu$ & $20.785\pm 0.033$ & $20.744\pm 0.018$ \\ 
$R_\tau$ & $20.764\pm 0.045$ & $20.790\pm 0.018$ \\ 
$R_b$ & $0.21638\pm 0.00066$ & $0.21569\pm 0.00016$ \\ 
$R_c$ & $0.1720\pm 0.0030$ & $0.17230\pm 0.00007$ \\ 
$Q_W(Cs)$ & $-72.65\pm 0.28\pm 0.34$ & $-73.10\pm 0.03$ \\
$Q_W(Tl)$ & $-116.6\pm 3.7$ & $-116.81\pm 0.04$ \\
$M_{Z_{1}}$(GeV) & $ 91.1872 \pm 0.0021 $ & $ 91.1870 \pm 0.0021 $ \\ 
\end{tabular}
\end{ruledtabular}
\end{table}

\begin{figure*}
\includegraphics{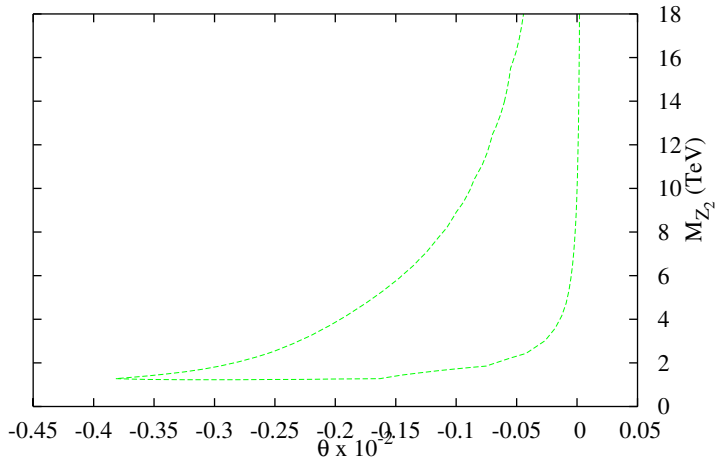}
\caption{\label{fig6}Contour plot displaying the allowed region for 
$\theta$ Vs. $M_{Z_2}$ at 95\% C.L..}
\end{figure*}

Introducing the expressions for $Z$ pole observable in Eq.(\ref{ancho}),
with $\Delta Q_W$ in terms of new physics in Eq.(\ref{DQ}) and using
experimental data from LEP, SLC and atomic parity violation (see Table
\ref{tab3}), we do a $\chi^2$ fit and we find the best allowed region in
the $(\theta-M_{Z_2})$ plane at $95\%$ confidence level (C.L.). In Fig.
\ref{fig6} we display this region which gives us the constraints
\begin{equation}\label{limits}
-0.0026\leq\theta\leq -0.0006, \;\;\; 2\; {\mbox TeV} \leq
M_{Z_2} \leq 100 \; {\mbox TeV},
\end{equation} 
with a central value at about 20 TeV.

As we can see the mass of the new neutral gauge boson is compatible with
the bound obtained in $p\bar{p}$ collisions at the Fermilab Tevatron
\cite{abe}. From our analysis we can see that  $M_{Z_2}$ peaks at a finite value larger than $100$~TeV
when for $\vert \theta \vert\rightarrow 0$, which still copes with the experimental constraints on the $\rho$
parameter.

\subsection{\label{sec:sec52}Bounds from unitary violation of the quark mixing matrix}
The see-saw mass mixing matrices for quarks and leptons presented in Eqs. (\ref{maup}), (\ref{madown}) and (\ref{malep}) are not a consequence of the particular discrete $Z_2$ symmetry introduced in Eq. (\ref{z2}); a stright forward calculation shows that any discrete  symmetry will reproduce the same mass matrices as far as we impose the following constraints:
\begin{itemize}
\item To have pure see-saw mass matrices in the Down, and Charged Lepton sectors.
\item To have a tree-level mass entry for the top quark mass in the third family, plus see-saw entries for the other two families in the Up quark sector.
\item To work with the non-minimal set of five Higgs scalars as introduced in the main text.
\end{itemize}
As a consequence of the mixing between ordinary and exotic quarks, violation of unitary in  the quark mixing matrix appears as discussed already in Sec.~\ref{sec:sec33}. This violation 
must be compatible with the experimental constraints of the mixing parameters as discussed in section 11 of Ref.~\cite{pdb}.

For the model discussed here, the structure of the quark mass matrices implies a mixing proportional to $\cos\delta$ (with $\delta = v/V$ as before) for the known quarks of each sector, which, when combined in the $V_{mix}$ entries, gives a mixing of the form $\cos^2\delta = 1-\sin^2\delta\approx 1-\delta^2$, being $\delta^2$ proportional to the violation of unitary in the model. Taking for $V\approx M_{Z_2}\approx 2$ TeV [the lower bound in Eq. (\ref{limits})], we obtain $\delta^2\approx 3.2\times 10^{-3}$, which is above the limit of the allowed unitary violation of $V_{mix}$~\cite{pdb}. As discussed in Sec.~\ref{sec:sec33}, a value of $V\approx 10$ TeV for the 3-3-1 mass scale is safe as far as the present violation of unitary in $V_{mix}$ is concerned.

\subsection{\label{sec:sec53}FCNC processes}
In a model like this, with four scalar triplets and mixing of ordinary with exotic fermion fields, we should worry about possible FCNC effects which may come either from the scalar sector, from the gauge boson sector, or from the unitary violation of $V_{mix}$.

First, notice that due to our $Z_2$ symmetry, FCNC coming from the scalar sector are not present at tree-level because each flavor couples only to a single scalar triplet. But FCNC effects can occur in $J_{\mu ,L}(Z)$ and $J_{\mu ,L}(Z^\prime)$ in Eqs. (\ref{jz1}) and (\ref{jz2}), respectively, due to the mixing of ordinary and heavy exotic fermion fields (notice from Eq. (\ref{jz1}) that $J_{\mu ,L}(Z)$ only includes as active quarks the three ordinary up and down-type quarks).

The stringest constraints in FCNC in the quark sector came from the transition $d\leftrightarrow s$, and the best place to look for them is in the $(K^0_L-K^0_S)$ mass difference, which may get contributions from the exchange of $Z^\mu_1$ and $Z^\mu_2$. The contribution from $Z^\mu_1$ is proportional to $|V^{*}_{us}V_{ud}|^2\approx |V^{*}_{us}V_{ud}|^2_{CKM} + 4\delta^4$ (where $|V^{*}_{us}V_{ud}|^2_{CKM}$ refers to the CKM). Then, the mixing of light and heavy quarks implies extra FCNC effects proportional to $4\delta^4$, which for $V\approx 2$ TeV implies a contribution to new FCNC effects above the allowed limits. So again, the 3-3-1 mass scale $V$ must be raised. Taking $V\approx 10$ TeV as discussed before, FCNC effects are now of the order of $10^{-7}$, value to be compared with the experimental bound $m(K_L)-m(K_s)\approx 3.48\pm 0.006\times 10^{-12}$ MeV \cite{pdb}, given that $4\delta^4< 0.006/3.48$, which means that in the context of this model there is room in the experimental uncertainties to include new FCNC effects, coming from the mixing between ordinary and exotic quarks.

Now, the FCNC contributions from $Z^\mu_2$ are safe, because they are not only constrained by the $\delta$ parameter, but also by the mixing angle $-0.0026\leq\theta\leq -0.0006$  as given in Eq.~(\ref{limits}).

\section{\label{sec:sec6}Conclusions}
During the last decade several 3-3-1 models for one~\cite{spm} and three families 
have been analyzed in the literature, the most popular one being the original 
Pisano-Pleitez-Frampton model~\cite{pf}. Other four different three-family 3-3-1 models are presented in Refs.~\cite{vl, ozer, sher}, one of them being the subject of study of this paper. The systematic analysis presented in
Refs.~\cite{pfs, pgs} shows that there are in fact an infinite number of
models based on the 3-3-1 local gauge structure, most of them including
particles with exotic electric charges. But the number of models with
particles without exotic electric charges are just a few~\cite{sher, pgs}.

In this paper we have carried out a systematic study of a 3-3-1 model that we have called a model with ``exotic charged leptons". In concrete, we have calculated for the first time its charged and neutral currents (see Tables \ref{tab1} and \ref{tab2}), we have embedded the structure into $SU(6)$ as a covering group, looked for unification possibilities, studied the gauge boson and fermion mass spectrum, and finally, by using a variety of experimental results, we have set constraints in several parameters of the model.

In our analysis we have done a detailed study of the conditions that produce a consistent charged fermion mass spectrum, a subject not even touched in the original paper~\cite{ozer}, except for a brief discussion of the neutrino sector done in Ref.~\cite{kita1}. First we have shown that a set of four Higgs scalars is enough to properly break the symmetry producing a consistent mass spectrum in the gauge boson sector. Then, the introduction of an appropriate anomaly-free discrete $Z_2$ symmetry plus an extra exotic down quark and a singlet scalar field, allow the construction of an appealing mass spectrum in the electrically charged fermion sector, without hierarchies in the Yukawa coupling constants. In particular we have carried a program for the quark sector in which: the four exotic quarks get heavy masses at the TeV scale, the top quark gets a tree-level mass at the electroweak scale, then the bottom, charm and strange quarks get see-saw masses and finally, the two quarks in the first family get radiative masses; the former without introducing strong hierarchies in the Yukawa coupling constants, neither new mass scales in the model.

The Higgs sector used in order to break the symmetry and to provide with masses to the charged fermions, plus whatever extra scalar fields could be needed to explain the masses and oscillations of the neutral lepton sector, renders the model with a quite complicated scalar potential, with several trilinear couplings possible (like for example $f\phi_1\phi_2\phi_3$ already used to give mass to the $u$ quark in the first family). This couplings are able to generate VEV for all the fields that feel them~\cite{fkl}. As a consequence, the pattern of spontaneous symmetry breaking becomes unstable and the minimization of the scalar potential may become a hopeless task. But this subject is far beyond the purpose of the present analysis.

We have also embedded the model into the covering group $SU(6)\supset SU(5)$ and studied the conditions for gauge coupling unification at a scale $M_G\approx 3\times 10^7$ GeV. The analysis has shown that a physical $(m_Z<M_V<M_G)$ one loop solution to the RGE can be achieved at the expense of introducing extra scalar fields at the intermediate energy scale $M_V$.

The fact that the RGE produces the same 3-3-1 mass scale than the lower limit obtained in the phenomenological analysis presented in Sec.\ref{sec:sec5} [compare Eqs. (\ref{scales}) and (\ref{limits})] is not accidental neither fortuitous. As a matter of fact, the extra scalar fields contributing to the beta functions in Eq.(\ref{betap331}), were just introduced for doing this job. A different set of scalar fields will produce either a different 3-3-1 and GUT mass scales, not unification at all, or either unphysical solutions. Eventhough our analysis may look a little arbitrary, we emphasize that we took the decision to play only with the most obscure part of any local gauge theory: the Higgs and scalar sectors.

Without looking at the neutral lepton sector, we may say that there are in this model only two mass scales: the 3-3-1 scale $V\geq 2$ TeV, and the electroweak scale $v\approx 10^2$ GeV.
Notice also that the discrete symmetry $Z_2$ introduced in the main text has the effect 
that each quark flavor gains a mass only from one Higgs field, which 
suppresses  possible FCNC effects.

What is lacking in this paper is a detailed analysis of the neutrino masses and oscillations. We could said that the study presented in Ref.~\cite{kita1} covers this part of the analysis; but unfortunately this is not the case. Comparing: the  authors in Ref.~\cite{kita1} use a different set of scalar fields, with a total set of just four scalar triplets, one of them being $\phi_5$ in Sec.~(\ref{sec:sec343}) which generates one-loop radiative Majorana masses, using the exotic heavy leptons as the seed; including also one electrically double charged Higgs scalar, singlet under $SU(3)_L$, which turns on the Zee-Babu mechanism. In their analysis they do not use a discrete symmetry, and they work under the assumption that the ordinary leptons ($e, \mu$ and $\tau$) has tree-level diagonal mass terms. Clearly, most of their assumptions do not fit in our picture. So, a detailed study of the neutral lepton sector must be done in the context of the model presented in this paper. Neutrino physics in this model is very rich and it deserves further attention.

Similar studies to the one presented here but for the model with ``right-handed neutrinos"~\cite{vl}, have been done in Refs.~\cite{canal, dwl}. Contrary to what is obtained here, the paper in Ref.~\cite{canal} shows that for the model with ``right-handed neutrinos", the see-saw mechanism for the Up and Down quark sectors can be implemented without including extra quark fields. But the model here does not need extra exotic electrons, which is the case for the model with ``right-handed neutrinos"~\cite{dwl}. Besides, the two models are embedded into $SU(6)$ as the common covering group, with extra scalar fields added in such a way that unification of the three gauge coupling constants is achieved at a relatively low energy scale, without conflict with proton decay bounds. Also, similar results for the bounds of the 3-3-1 mass scale $V$ and mixing angle $\theta$~\cite{dwl} were found.

We have presented in this paper, original results compared with previous 
analysis~\cite{ozer, kita1}. First and most important, our Higgs sector and VEV are 
different to the ones introduced in the original paper~\cite{ozer}. They imply 
different mass matrices for gauge bosons and fermion fields, with quite a different phenomenology. The most important fact about our Higgs sector is that it allows for a consistent charged fermion mass spectrum, without a strong hierarchy between the Yukawa coupling constants. Besides, it allows for the first time in the context of the model, the identification of the quark mass eigenstates, as a function of the weak states. Using that information, a consistent phenomenologycal analysis which sets reliable bounds on new physics coming from heavy neutral currents can be done.

As far as the particle spectrum is concerned, let us say that in the scalar sector, and according to the ESH~\cite{esh}, at least one more $SU(2)_L$ neutral singlet and a second Higgs doublet should show up at the electroweak scale, with all the other Higgs scalars getting a mass at the TeV scale (the neutral singlet does not couple to the SM fermions at the tree level). For the charged fermions, the four exotic quarks (two Up and two Down) and the three exotic electrons should get masses at the 3-3-1 scale (2 TeV $\leq V\leq $10 TeV). Some of these particles should show up at the forthcoming LHC facilities.

\section*{Acknowledgments}

We acknowledge partial financial support from COLCIENCIAS in Colombia, and from CODI in the ``Universidad de Antioquia"

\vspace{.5cm}

\textit{Note added in proof}: During the time period of revision of the original manuscript, the paper: ``Lepton masses and mixing without Yukawa Hierarchies" by W.A.Ponce and O.Zapata, appears~\cite{ponza}. That paper addresses, in full detail, the lepton sector of the model studied here; in particular, a set of constraints were impposed in the parameters of the lepton sector such that lepton FCNC and neutrinoless double beta decay, became suppressed below the experimental bounds.

\end{document}